\documentclass[pss]{wiley2sp} 
\usepackage{amsmath}
\usepackage{amsbsy}

\usepackage[colorlinks=true,citecolor=blue,linkcolor=red,linktocpage=true,pagebackref=false]{hyperref}
\usepackage{soul}
\usepackage{color}
\usepackage{varwidth}
\usepackage{times}
\usepackage{paralist}
\usepackage{textcomp} 
\usepackage{changes}

\definecolor{geraldine}{rgb}{1.0,0.5,0.0}
\definecolor{misha}{rgb}{0.0,0.7,0.7}

\begin{document}



\title{Heat and charge transport measurements to access single-electron quantum characteristics}

\titlerunning{Transport properties of single-electron excitations}

\author{%
  Michael Moskalets\textsuperscript{\Ast,\textsf{\bfseries 1}},
  G\'eraldine Haack\textsuperscript{\textsf{\bfseries 2}}
  }

\authorrunning{M. Moskalets and G. Haack}

\mail{e-mail
  \textsf{michael.moskalets@gmail.com}, 
  Phone: +380-57-707-68-31, 
  Fax: +380-57-707-66-01 
  }

\institute{%
  \textsuperscript{1}\,Department of Metal and Semiconductor Physics, NTU ``Kharkiv Polytechnic Institute", 61002 Kharkiv, Ukraine\\
  \textsuperscript{2}\,D\'epartement de Physique Th\'eorique, Universit\'e de Gen\`eve, CH-1211 Gen\`eve 4, Switzerland
  }

\received{XXXX, revised XXXX, accepted XXXX} 
\published{XXXX} 


\abstract{%
%
%
%
\abstcol{
In the framework of the Floquet scattering-matrix theory we discuss how electrical and heat currents accessible in mesoscopics are related to the state of excitations injected by a single-electron source into an electron waveguide. 
We put forward an interpretation of a single-particle heat current, which differs essentially from that of an electrical current. 
We show that the knowledge of both a time-dependent electrical current  and a time-dependent heat current allows the full reconstruction of a single-electron wave function.
}{
In addition we compare electrical and heat shot noise caused by splitting of a regular stream of single-electron excitations. If only one stream impinges on a wave splitter, the heat shot noise is proportional to the well-known expression of the charge shot noise, reflecting the partitioning of the incoming single particles. The situation differs when two electronic streams collide at the wave splitter. The shot noise suppression, due to the Pauli exclusion principle, is governed by different overlap integrals in the case of charge and of heat.}
}


\maketitle   


\section{Introduction}

The experimental realization of an on-demand high speed single-electron source \cite{Blumenthal:2007ho,Feve:2007jx,Dubois:2013dv} is a major step on the way of implementing of a fermionic platform for quantum information processing. 
Such a platform potentially provides a high level of miniaturization, is scalable, and takes advantage of the industrial planar technology. 

Single electrons are the most compact carrier of information in solid-state 
ballistic conductors. 
As it was shown experimentally, electrons can be transferred on demand between distant quantum dots one by one.\cite{Hermelin:2011du,McNeil:2011ex}
Moreover the flying qubit with electrons was already reported.\cite{Yamamoto:2012bp}
Ideally information is encoded into the state of a single particle, in its wave function. 
The way to acquire this information, which is natural for solid-state mesoscopic systems, is to perform an appropriate transport measurement. 

In this work, we aim at providing general analytical expressions for transport characteristics of a multi-terminal mesoscopic 
conductor,
in terms of electronic correlation functions, and more specifically for few examples of state-of-the-art single-electron states. These expressions and their understanding bring fundamental additional information about single-electron states and allow us to propose alternative measurements to perform tomography of single-particle states. 

In particular, we show that the wave-function of single-electronic excitations can be reconstructed from time-dependent charge and heat currents. 
Based on the analytical expressions we derive for these two quantities 
within the Floquet scattering matrix formalism \cite{Moskalets:2011cw}, 
we define the energy per emitted particle. Its non-monotonous behavior as a function of time reflects the different interpretation we make of time-dependent charge and heat currents and might be exploited as a resource for thermodynamical tasks.

\section{Single-electron excitations}

The state of injected electrons 
depends on the type of source and its working regime. 
Generally the state emitted at zero ambient temperature is pure and it can be characterized by a wave function.\cite{Moskalets:2016dx} 
For illustrative purposes, we present below three known analytical expression for a wave function of a single-electron injected on the top of the Fermi sea in different regimes. 

{\it First example}: 
Driving a quantum capacitor,\cite{Buttiker:1993wh,Gabelli:2006eg} which is a chiral one-dimensional quantum dot, by a step potential at optimal conditions \cite{Feve:2007jx} leads to the emission of an electron with the following wave function, $\Psi_{Tr}(t) = e^{-i t \frac{ \mu }{ \hbar} } \psi_{Tr}(t-t_{e})$, \cite{Moskalets:2013dl,Haack:2013ch}. Here $t_{e}$ is the time of emission, \textit{i.e.} the time when a quantum level suddenly arises above the Fermi level with energy $\mu$, and 
\begin{eqnarray}
 \psi_{Tr}(t) = \frac{1 }{ \sqrt{v_{ \mu}} } e^{-i t \frac{ \Delta   }{ 2 \hbar } }  \theta(t) \frac{ e^{ -\frac{t }{2 \tau_{D} }  } }{ \sqrt{\tau_{D} } }  . 
\label{Tr-wf}
\end{eqnarray}
Here $v_{ \mu}$ is the Fermi velocity.
An electron is emitted during a transient process (hence a subscript "Tr") of duration $\tau_{D}$. 
The optimal conditions imply the following: (i) There is an equidistant ladder of levels in a quantum dot around the Fermi energy; (ii) The Fermi level is positioned exactly in the middle between the two subsequent levels; (iii) The potential applied to the dot shifts all the ladder by one level spacing $\Delta$. 
As a result a single occupied level suddenly raises above the Fermi level with an excess energy $\Delta/2$. 

We use a convention that a wave function $\Psi(t)$ is given as a function of time just behind the source and it is normalized as $ \int dt \left | \Psi(t) \right |^{2} = 1/v_{ \mu}$. 
Such a convention is natural for mesoscopics, where electron detectors are   fixed in space rather than in time. 
To calculate the wave function at a distance $x$ away from the source, one needs to shift the time as following:  $t \to t - x/v_{\mu}$ and $t \mu \to t \mu - p_{ \mu} x$, where $p _{\mu} = \sqrt{2 m \mu}$ is a momentum  of electrons with mass $m$. 
In the coordinate space, the normalization condition takes its ordinary form, $\int dx \left | \Psi(t - x/v_{\mu}) \right |^{2} = 1$.

{\it Second example}: 
We consider a voltage pulse, uniform in space and characterized by a Lorentzian shape in time and a unit flux of the form: $eV(t) = 2 \hbar \Gamma_{\tau}/\left( \left[ t - t_e \right]^{2}+ \Gamma_{\tau}^2  \right)$.
Applying this voltage pulse to the Fermi sea, a single electron is created \cite{Levitov:1996ie,Ivanov:1997kf} with a wave function $\Psi_{L}(t) = e^{-i t \frac{ \mu }{ \hbar} } \psi_{L}(t-t_{e})$, \cite{{Keeling:2006hq},{Dubois:2013fs}} where $t_{e}$ is the time of emission, the time when a voltage pulse has a maximum, and the envelop function is given by:
\begin{eqnarray}
 \psi_{L}(t) =      \frac{1 }{ \sqrt{v_{ \mu}} } \sqrt{\frac{1}{  \pi \Gamma _{\tau}  } }\frac{ 1 }{t/  \Gamma _{\tau} - i } \,.
\label{L-wf}
\end{eqnarray}
Here $\Gamma_{\tau}$ is the half-width of the voltage pulse. 
Such a particle was observed experimentally and named {\it a leviton} \cite{Dubois:2013dv}, hence the subscript "L".
The wave function of a leviton was directly measured in Ref.~\cite{Jullien:2014ii}.  

Let us remark that an electron with the same wave function as that of a leviton \cite{Moskalets:2013dl,Haack:2013ch} can be emitted by a quantum capacitor,  the experimental setup used in the first example mentioned above.
For this purpose, the  driving potential should vary slowly in time, such that  the time interval $2 \Gamma_{\tau}$ during which a rising level crosses the Fermi level would be large compared to the dwell time. 
This dwell time corresponds to the time during which an electron leaves  the  capacitor if there are empty states outside.\cite{Splettstoesser:2008gc} 
Such a regime, namely when $ \Gamma_{\tau} \gg \tau_{D}$,  is referred to as {\it an adiabatic regime of emission}.
If the crossing time $2 \Gamma_{\tau}$ becomes comparable to the dwell time $\tau_{D }$, the process of emission becomes non-adiabatic.\cite{Moskalets:2013dl}  

{\it Third example}:
The wave function valid in both adiabatic and non-adiabatic regimes was found  for the case of a quantum level raising above the Fermi level  at a  constant 
rapidity  
$c$.\cite{Keeling:2008ft}.
We mark the corresponding quantities with the subscript "CS". 
The wave function is $\Psi_{CS}(t) = e^{-i t \frac{ \mu }{ \hbar} } \psi_{CS}(t-t_{e})$, where 
\begin{eqnarray}
 \psi_{CS}(t) &=& 
 \frac{1 }{ \sqrt{v_{ \mu}} }
 \frac{ 1 }{\sqrt{ \pi  \Gamma _{\tau} }}  
 \int _{0}^{ \infty } \frac{d \epsilon }{ 2\epsilon_{0} } 
\nonumber \\ 
\label{CS-wf}\\
&& \times
\exp\left\{ -i \frac{\epsilon }{ 2 \epsilon_{0} } \frac{t }{ \Gamma _{\tau} } - \frac{ \epsilon }{2 \epsilon_{0} } + i \zeta \left( \frac{ \epsilon }{ 2\epsilon_{0} }  \right)^{2} \right\} .
 \nonumber 
\end{eqnarray}
Here $ \epsilon_{0} = \hbar/  (2\Gamma _{\tau})$ corresponds to the energy of an excitation. It can also be expressed in terms of the dwell time, $\epsilon_0 = c \tau_D$. The parameter $\zeta= \epsilon_{0}/ \gamma$ is dimensionless and $\gamma$ is the level width.

Since the crossing time is inversely proportional to the level's 
rapidity, 
$ \Gamma_{\tau} = \gamma / c$, we see that  $\zeta \sim c$. 
In the limit when the level rises slowly, $c\to 0$, we can put $ \zeta=0$ and  Eq.~(\ref{CS-wf}) reproduces Eq.~(\ref{L-wf}).  
In this case the extension of the single-particle wave function is defined by the crossing time $ \Gamma_{\tau}$, which increases with decreasing 
rapidity 
$c$. 
In contrast, in the limit $c\to \infty$, despite the fact that the crossing time $ \Gamma_{\tau}\to 0$, the wave packet does not shrink down to zero  since an electron escapes into the Fermi sea during a finite time independent of $c$, namely the dwell time, $ \tau_{D} = \hbar/(2 \gamma)$.  
In this case the density profile $\left |  \Psi_{CS}(t) \right |^{2}$ resembles the density profile $\left | \Psi_{Tr}(t) \right |^{2}$, see Eq.~(\ref{Tr-wf}). 
However the energy properties of an excitation with a wave function $ \Psi_{CS}$ are quite different from the ones of an excitation described by a wave function $ \Psi_{Tr}$.\cite{Pascal} 
In particular, the energy distribution of the former excitation is exponential with  mean energy $ \epsilon_{0}$ increasing with $c$,  while the energy distribution of the latter one is Lorentzian with constant mean energy $ \Delta/2$. 
Let us remark that the energy distribution for particles with wave functions from the second and the third examples is identical as far as the parameter $ \Gamma_{\tau}$ is the same. 

As it was shown in Ref.~\cite{Moskalets:2015kx}, for an adiabatically created one-dimensional spinless excitation, the density profile $\left |  \Psi(t) \right |^{2}$ uniquely defines its wave function. 
Therefore, a measurement of a time-dependent electrical current is enough to determine an electron wave function.
This is no longer the case for non-adiabatically created excitations. 
For instance, the wave functions $\Psi_{Tr}$ and $ \Psi_{CS}$ at $c\to \infty$ are different, even though their squares (density profiles) are the same.   
As we show below, an electrical current and a heat current together are sufficient to define a wave function of excitations created arbitrary, adiabatically or non-adiabatically.

\section{A time-dependent current of a single-electron excitation}

In a wide-band approximation, the electrical current $I(t)$ and the heat current $I^{Q}(t)$ associated to a single-particle excitation with a wave function $\Psi(t) =e^{-i t \frac{ \mu }{ \hbar} } \psi(t)$  just behind the source are calculated as follows:
\begin{subequations}
\label{curr}
\begin{eqnarray}
I(t) = v_{ \mu} e  \left | \psi(t) \right |^{2}, 
\label{curra}
\end{eqnarray}
\begin{eqnarray}
I^{Q}(t) = v_{ \mu} \hbar \, {\rm Im} \left [ \frac{ \partial \psi^{*}(t) }{ \partial t } \psi(t) \right] .
\label{currb}
\end{eqnarray}
\end{subequations}
Here ${\rm Im} \left[ X \right]$ stays for the  imaginary part of $X$. The positive direction for both currents is defined from the source into the electronic waveguide. 
We refer to Append.~\ref{appa}, Append.~\ref{eFl}, and Append.~\ref{hFl} 
for relations valid  for multi-particle excitations. 

Equations \ref{curra} and \ref{currb} indicate that time-dependent electrical and heat (energy) currents are fundamentally different, \cite{Ludovico:2014de,Rossello:2015iw,Calzona16}, already at a single-particle level. 
Below, we investigate and exploit these differences.

\subsection{Quantum-mechanical analogy}

Equations \ref{curr}  were derived within the Floquet scattering-matrix approach for the regime of a single-particle emission. 
They are also in full agreement with the conventional quantum-mechanical expression. 
Indeed the electrical current $I(t)$ is defined as an electron charge $e$ multiplied by  a probability current $- \frac{ \hbar }{ m }  {\rm Im} \left [ \frac{ \partial \Psi^{*} }{ \partial x } \Psi \right] $.
Using $ \Psi(x,t) = e^{-i  \frac{ t \mu - p_{ \mu} x }{ \hbar  } } \psi\left( t - \frac{ x }{ v_{ \mu} }  \right)$ and  the wide-band approximation, we can neglect the spatial variation of the envelope function $\partial \psi/\partial x$ compared to the inverse Fermi wave-length $p_{ \mu}/ \hbar$, which leads to Eq.~(\ref{curra}).

In the same way the heat current $I^{Q}(t) = I^{E}(t) - \mu I(t)/e$, Eq.~(\ref{currb}), can be calculated using a less wide-known quantum mechanical equation of an energy current for a particle with mass $m$ and with squared dispersion relation, $I^{E}(t) = \frac{\hbar^{2}}{2m} {\rm Re} \left[  \Psi^{*}  \frac{\partial^{2}  \Psi }{\partial t\partial x}   - \frac{\partial  \Psi^{*}  }{\partial x} \frac{\partial  \Psi  }{\partial t} \right] $, where ${\rm Re} \left[ X \right]$ is the  real part of $X$.\cite{Mathews:1974kb,Levin:2012ki,Ludovico:2014de}

\subsection{Interpretation of a time-dependent current of a single particle}

It is well-known that an electrical (energy) current is defined as an amount of charge (energy) transferred per unit of time. 
However, if we consider a current carried by a single electron, such an interpretation is not applicable since an electron is not divisible and can be detected only entirely. 

As it follows from Eq.~(\ref{curra}), an electrical current of an electron emitted by the source is interpreted in the same way as (the modulus squared of) a wave function: A current at a time $t$ is given by the probability density to detect an electron at that time multiplied by an electron charge. 
So, if we take an ensemble of identical particles and measure a time-resolved detection statistics, we would then obtain a time-dependent electrical current. 
A periodically working source emits  an ensemble of identical particles and is therefore  suitable for this purpose. 
This was confirmed experimentally when a time-dependent current of an electron with wave function $ \Psi_{Tr}$, Eq.~(\ref{Tr-wf}) was measured in Ref.~\cite{Feve:2007jx} using a periodically-working source. 
Let us emphasize that the interpretation of an electrical current $I(t)$ as a probability current, Eq.~(\ref{curra}), is mainly based on the fact that an electron charge is indivisible, \textit{i.e.} it can only be measured entirely. However, the probability to detect an electron charge is varying in time, according to the density profile.

In the case of a time-dependent heat current $I^{Q}(t)$, Eq.~(\ref{currb}), the interpretation is more subtle as it requires the analysis of the detection process. 
One can conjecture that the amount of energy by the single-electron excitation depends on the time at which the particle is detected. 
We therefore distinguish two regions behind the source: (i) {\it the near field region}, when the distance to the source is shorter than the spatial extension of a single-particle excitation, and (ii) {\it the far field region}, when the distance to the source exceeds the size of a wave packet. 

If the detector is located in the far field region, the emission process is completed \textit{before} the time of detection. 
Hence,  the detected particle carries a fixed amount of energy which is independent of a precise time of detection. 
The detection process can be made for instance with a quantum dot acting as energy filter. \cite{Lee:2013fc,Jezouin:2013fx,Gasparinetti:2015gr}. 
This energy, defined as the integral over time of a heat current, is what is usually understood as energy of an emitted particle (counted from the Fermi energy), 
\begin{eqnarray}
Q = \int_{- \infty}^{ \infty} dt^{\prime} I^{Q}(t^{\prime}) .
\label{Qinf}
\end{eqnarray}
For the three examples presented in the introduction, the energy is given respectively by $Q_{Tr} = \Delta/2$ \cite{Moskalets:2013dl},   $Q_{L} = \hbar/(2 \Gamma_{\tau})$ \cite{Keeling:2006hq,{Moskalets:2009dk}} and $Q_{CS} = \hbar/(2 \Gamma_{\tau})$.   

If the detector is located in the near field region,  the emission process is not yet completed when a particle is detected and the energy of a particle therefore depends on the time of detection.   
One can say that what is detected in the near field region is the energy which has flowed between the source and the detector up to a detection time $t$. 
This energy is defined as $ \delta Q = \int_{- \infty}^{t} dt^{\prime} I^{Q}(t^{\prime})$.  
To calculate the energy per particle, we need to take into account that, formally, the amount of energy $ \delta Q$ is carried on average by the  number of particles $ \delta N = \int_{- \infty}^{t} dt^{\prime} I(t^{\prime})$. 
This number is smaller than one since the probability to detect a particle (even for an ideal detector) is smaller than one in the near field region. 
Therefore, the energy per detected particle should be enhanced compared to $ \delta Q$ by a factor $1/ \delta N$: 
\begin{eqnarray}
Q(t) =\frac{ \int_{ - \infty}^{t} dt^{\prime} I^{Q}(t^{\prime}) }{  \int_{- \infty}^{t} dt^{\prime} I(t^{\prime})/e } .
\label{Qt}
\end{eqnarray}
When the emission process is completed, $t \to\infty$, the equation above agrees with Eq.~(\ref{Qinf}), $Q( \infty) = Q$, since the total probability to emit a particle is one, $ \int_{- \infty}^{ \infty} dt^{\prime} I(t^{\prime})/e=1$.

Equation 6 reflects the interpretation we put forward in this work for the energy detected per particle at a given time $t$. It corresponds to the integrated heat current until the time of detection $t$, weighted by the corresponding number of emitted particle until this same time $t$.

As we already mentioned, the size $ \lambda_{nf}$ of the near field region is  set  by the spatial extension of a single-electron wave packet. 
For instance, the duration of a leviton from Ref.~\cite{Dubois:2013dv} is $  \Gamma _{\tau} \sim 3 \times 10^{-11}\ s$. 
Using a characteristic Fermi velocity $v_{ \mu} \sim 10^{5} \div 10^{6}\ m/s$ \cite{{Kumada:2014fl},Kataoka:2016cv} we get $ \lambda_{nf} \sim  \Gamma _{\tau}\times v_{ \mu} = 3 \div 30 \mu m$. 
The characteristic size of a mesoscopic electronic 
conductor 
being of the order of  a few microns,  the measurement of a time-dependent heat current is within reach given the state-of-the-art of the experiments.

We stress that $Q(t)$ is a particle's energy understood as a quantum-mechanical average. 
Let us remark that fluctuations of this energy  are caused by the probabilistic nature of interaction between a dynamical source and electrons as discussed in Refs \cite{Battista:2013ew,Moskalets:2014ea,Moskalets:2014cr,{Battista:2014di}}.  

To clarify further the interpretation presented above, let us analyze the relation between the electrical and heat currents using the analytical expressions of the three electron wave functions taken as examples.

\subsubsection{First example}
\label{1eQI}

Using the wave function $ \Psi_{Tr}$, Eq.~(\ref{Tr-wf}), in Eqs.~(\ref{curr}) we get a time-dependent heat current,
\begin{eqnarray}
I_{Tr}^{Q}(t) &=& \theta(t) \frac{  \Delta }{ 2 \tau_{D} } e^{ -\frac{t }{ \tau_{D} }  } 
\nonumber \\
\label{Tr-Q-e} \\
&=& 
\frac{ \Delta }{ 2 } \frac{I_{Tr}(t) }{ e }, 
\nonumber 
\end{eqnarray}
where $I_{tr}(t) =  \theta(t) \frac{ e }{ \tau_{D} } e^{ -\frac{t }{ \tau_{D} }  }$ \cite{Feve:2007jx,{Moskalets:2008fz}} is an electrical current.  
The linear dependence between the charge current and the heat current is characteristic of a spontaneous decaying process, since there is no external source either changing the energy of a particle during its decay or able to affect the rate of the decay process. 
The quantum level is raised by $ \Delta/2$ over the Fermi level and, therefore, an emitted electron carries a fixed amount of an excess energy (heat), $ \Delta/2$, no matter when it is actually emitted or detected. 
The use of Eqs.~(\ref{Qinf}) and (\ref{Qt}) gives
\begin{eqnarray}
Q_{Tr}(t) = Q_{Tr} = \Delta/2.
\end{eqnarray}

The situation is radically different when the emission process does not correspond to a spontaneous decay. In this case,  the energy of an emitted particle does depend on an actual time of emission and detection as it occurs in the next examples. 

\subsubsection{Second example}

Substituting the wave function $\Psi_{L}$, Eq.~(\ref{L-wf}) into Eqs.~(\ref{curr}), we find that the heat and charge currents of a leviton obey the Joule law, 
\begin{eqnarray}
 I_{L}^{Q}(t) &=& \frac{ 2 \epsilon_{0}  }{ \pi  \Gamma _{\tau} }  \frac{ 1 }{\left [ (t/  \Gamma _{\tau})^{2} +  1 \right]^{2}  }
\nonumber \\ 
\label{Jlaw} \\
 &=& R_{q} I_{L}^{2}(t) ,
 \nonumber 
\end{eqnarray}
where $I_{L}(t) =  \frac{ e }{ \pi  \Gamma _{\tau} }  \frac{ 1 }{(t/  \Gamma _{\tau})^{2} +  1  }$ \cite{{Levitov:1996ie},{Ivanov:1997kf},{Keeling:2006hq},{Dubois:2013fs}} is a time-dependent current of a leviton and $R_{q}= h/(2e^{2})$ is the B\"{u}ttiker resistance, also known as the charge relaxation resistance quantum \cite{Buttiker:1993wh}. 
 
Note that the Joule law is expected to hold in a one-dimensional conductor with non-interacting electrons at arbitrary but adiabatic driving.
\cite{Ludovico:2014de,Ludovico:2016bo}
Here we demonstrate that it holds even on a single-particle level. 

In macroscopic conductors, the quadratic dependence of the heat current with the charge current can be understood as follows. 
The driving force (say, a voltage across a conductor) defines on one hand the rate of particle transfer (an electrical current) and, on the other hand, the energy acquired by the flowing  particles (the voltage drop). 
The heat current, corresponding to the rate of released heat, is the product of the rate of a particle transfer and the energy per particle and is therefore quadratic in an external potential or, equivalently, in a charge current. 

\begin{figure}[b]
\centerline{
\includegraphics[width=80mm]{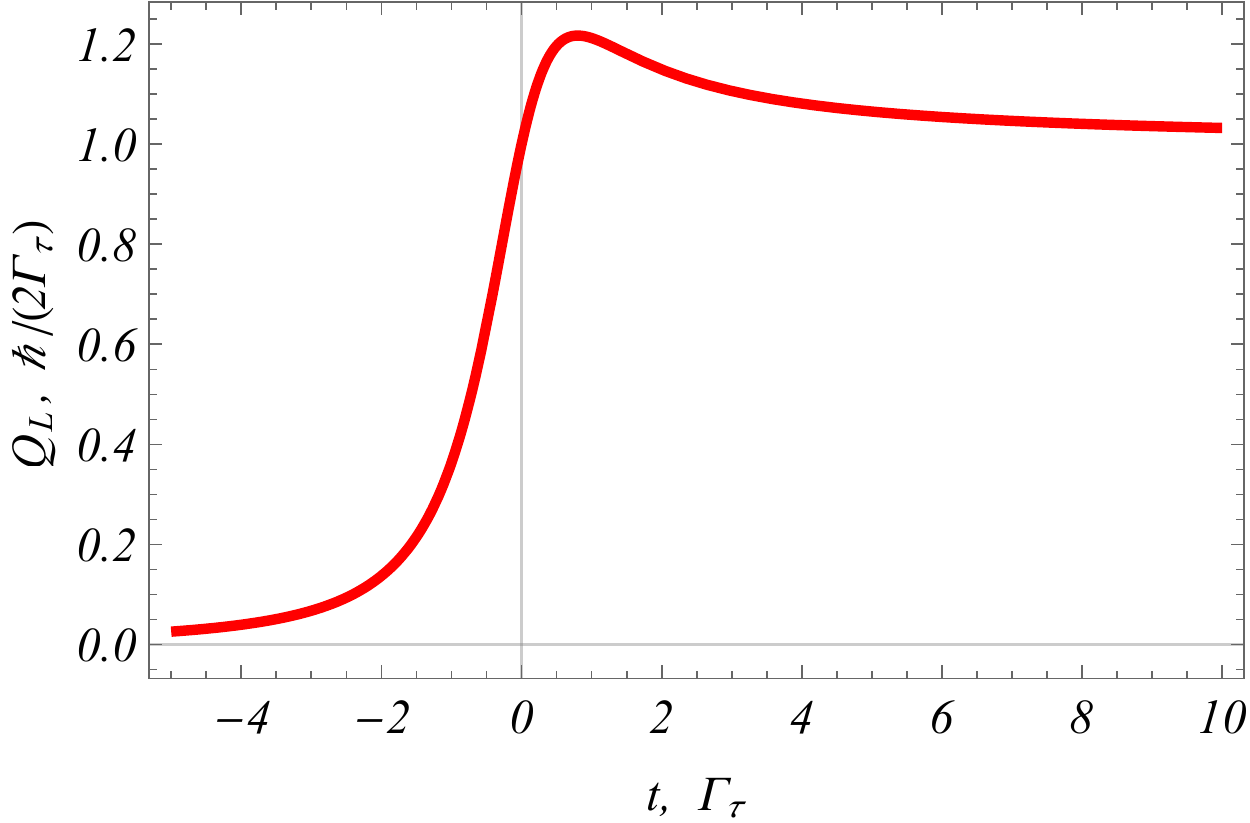}
}
\caption{(Color online) 
A time-dependent energy per leviton $ Q_{L}(t)$, see Eq.~(\ref{Qt}) calculated for $ \Psi_{L}(t)$ from Eq.~(\ref{L-wf}). 
The energy is normalized to $ Q_{L}( \infty) = \hbar/(2  \Gamma _{\tau})$
The time of emission is $t_{e} = 0$. 
 }
\label{ELt}
\end{figure}

This reasoning suggests that the energy of a leviton depends on the detection's time. 
Indeed, the leviton gets its energy from a time dependent potential  which creates it: The longer a leviton is in touch with a parental field, the larger its energy becomes. 
However, in fact, the energy of a leviton detected in the near field region, Eq.~(\ref{Qt}), is a non-monotonous function of time; it has a slight maximum near the time of emission, see Fig.~\ref{ELt}.  

\begin{figure*}[t]
\centerline{
\includegraphics[width=0.85\textwidth]{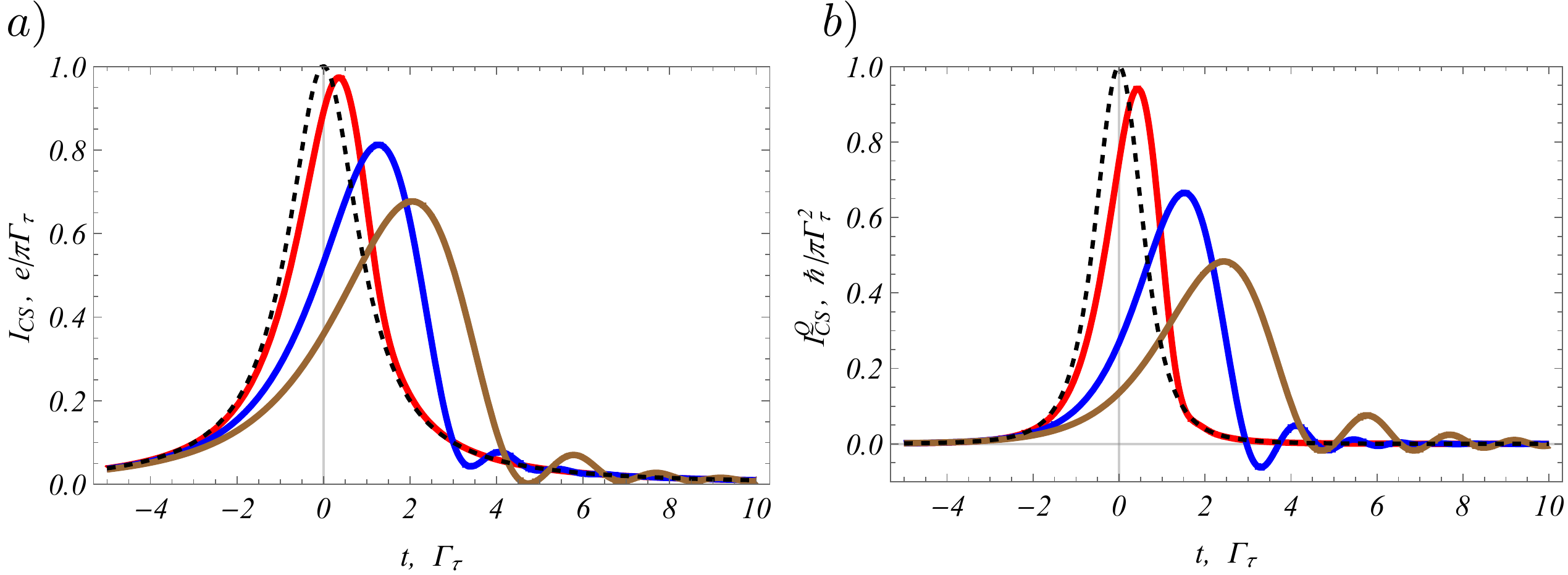}
}
\caption{(Color online) 
a) Time-dependent electrical current $I_{CS}(t)$, Eq.~(\ref{CScurra}), and b) Time-dependent heat current $I_{CS}^{Q}(t)$, Eq.~(\ref{CScurrb}). 
The solid lines are for $ \zeta = 0.1, 0.5, 1$ in the order of decaying amplitudes and the time of emission is set to $t_{e} = 0$. 
A dashed line,  $ \zeta = 0$, reproduces on panel a) the current carried by a Leviton, $I_{L}(t) =  \frac{ e }{ \pi  \Gamma _{\tau} }  \frac{ 1 }{(t/  \Gamma _{\tau})^{2} +  1  }$ and on panel b) the heat current carried by a Leviton, $I_{L}^{Q}(t) =  \frac{ 2 \epsilon_{0}  }{ \pi  \Gamma _{\tau} }  \frac{ 1 }{\left [ (t/  \Gamma _{\tau})^{2} +  1 \right]^{2}  }$, Eq.~(\ref{Jlaw}).
}
\label{ICS}
\end{figure*}

Formally this non-monotonous behaviour is caused by the fact that $I_{L}^{Q}(t)$ has a narrower peak compared to that of $I_{L}(t)$, compare dashed lines in Fig.~\ref{ICS}, Panels a) and b). 
Physically a non-monotonous behaviour of $ Q_{L}(t)$ is a manifestation of a quantum-coherent evolution of a single-particle state in time during its emission/creation. This evolution is governed by interferences of amplitudes corresponding to  the interaction between a particle and a time-dependent field driving the source at different times.\cite{Moskalets:2008ii}

\subsubsection{Third example}

An electron with wave function $ \Psi_{CS}$, Eq.~(\ref{CS-wf}), carries   a charge  and a heat currents given by 

\begin{subequations}
\label{CScurr}
\begin{eqnarray}
I_{CS}(t) &=& 
\frac{ e }{ \pi  \Gamma _{\tau} } 
 \iint _{0}^{ \infty } \frac{d \epsilon }{ 2 \epsilon_{0} } \frac{d \epsilon^{\prime} }{ 2 \epsilon_{0} } e^{ - \frac{ \epsilon + \epsilon^{\prime}  }{2 \epsilon_{0} }  } 
\nonumber \\
\label{CScurra} \\
&&\times
\cos  \left[  \frac{\epsilon^{\prime} - \epsilon }{ 2 \epsilon_{0} } \frac{ t }{  \Gamma _{\tau} } +  \zeta \frac{ \epsilon^{2} - \left( \epsilon^{\prime}  \right)^{2} }{4 \epsilon_{0}^{2} }  \right] .
\nonumber 
\end{eqnarray}
\begin{eqnarray}
I_{CS}^{Q}(t) &=& 
\frac{ 2 \epsilon_{0} }{ \pi  \Gamma _{\tau} } 
\iint _{0}^{ \infty } \frac{d \epsilon }{ 2 \epsilon_{0} } \frac{d \epsilon^{\prime} }{ 2 \epsilon_{0} } 
\frac{ \epsilon^{\prime} }{ 2 \epsilon_{0} } 
e^{ - \frac{ \epsilon + \epsilon^{\prime}  }{2 \epsilon_{0} }  } 
\nonumber \\
\label{CScurrb} \\
&&\times
\cos  \left[  \frac{\epsilon^{\prime} - \epsilon }{ 2 \epsilon_{0} } \frac{ t }{  \Gamma _{\tau} } +  \zeta \frac{ \epsilon^{2} - \left( \epsilon^{\prime}  \right)^{2} }{4 \epsilon_{0}^{2} }  \right] .
\nonumber 
\end{eqnarray}
\end{subequations}
\ \\ \noindent
At non-zero (and not too large) $ \zeta$, both currents in Fig.~\ref{ICS} 
exhibit oscillations.



Interestingly, in the long time limit, the energy of an emitted particle is independent of the parameter $\zeta$ and it coincides with the energy of a leviton having the same parameter $ \Gamma_{\tau}$:   $ Q_{CS} = \int_{ - \infty}^{ \infty} dt^{\prime} I_{CS}^{Q}(t^{\prime}) = \hbar /(2  \Gamma _{\tau})$.

The line corresponding to $ \zeta=0.5$ on 
Fig.~\ref{ICS} b) clearly demonstrates a negative heat flux around $t=3.5  \Gamma _{\tau}$. 
We stress that such a negative flux does not mean that  heat flows into a rising quantum level from a zero-temperature Fermi sea, what would be unphysical. 
According to Eq.~(\ref{Qt}), a negative heat flux simply means that the energy of a particle detected earlier can be larger than the energy of a particle detected later. 

Note that a negative time-resolved heat flux was previously reported in Ref.~\cite{Ludovico:2014gq} for a fast driven resonant level system coupled to a zero-temperature fermionic reservoir.

\begin{figure}[b]
\centerline{
\includegraphics[width=80mm]{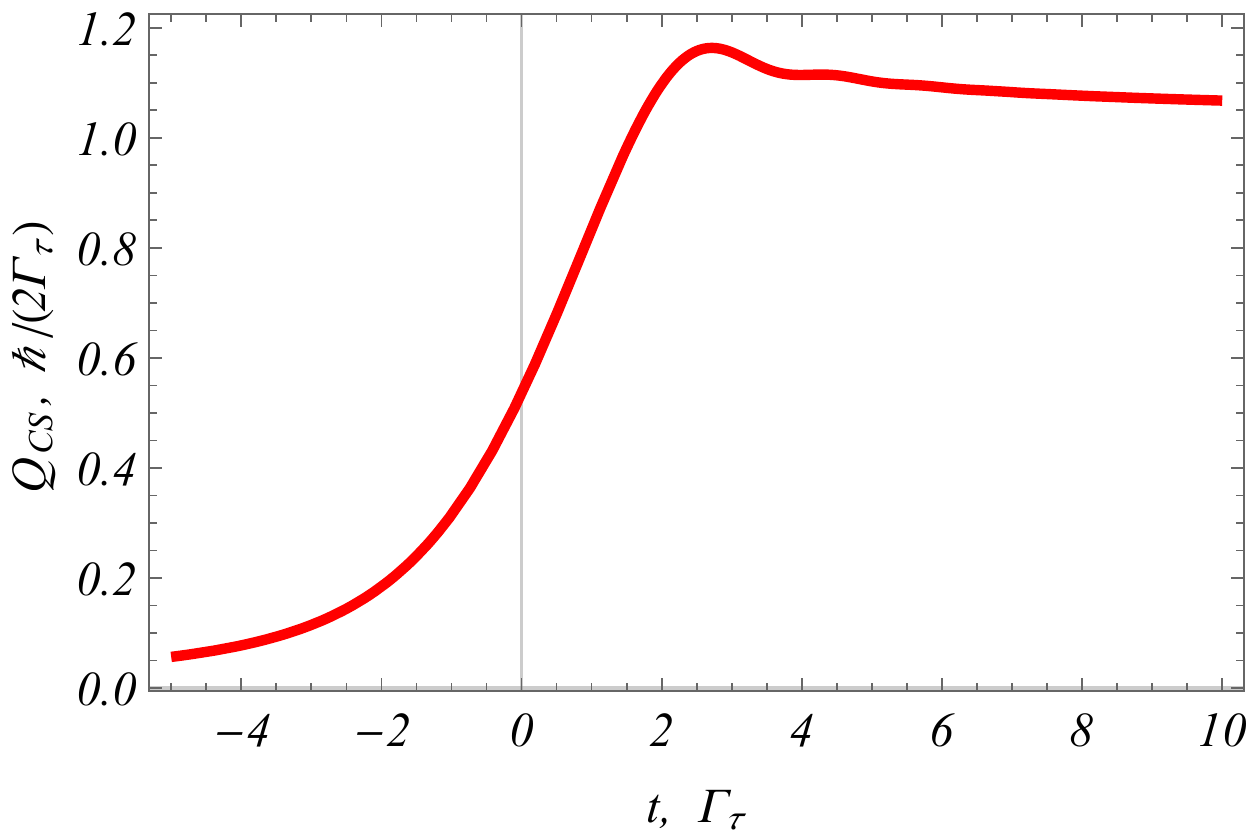}
}
\caption{(Color online) 
A time-dependent energy per particle emitted from the level raising 
at a constant rapidity 
$ Q_{CS}(t)$, see Eq.~(\ref{Qt}) calculated for $ \Psi_{CS}(t)$ from Eq.~(\ref{CS-wf}) with parameter $ \zeta = 0.5$. 
The energy is normalized to $ Q_{CS}( \infty) = \hbar/(2  \Gamma _{\tau})$
The time of emission is $t_{e} = 0$. 
 }
\label{ECSt}
\end{figure}

In Fig.~\ref{ECSt} we show a time-dependent energy for a particle described by $ \Psi_{CS}$. 
We see that, as in the case of a leviton, this energy has a maximum at some intermediate time. 
Therefore, by absorbing an electron earlier, one can extract more energy from the electronic source. 

We stress that the maxima at intermediate time in Fig.~\ref{ECSt} is still present when the energy $ \delta Q = \int_{- \infty}^{t} dt^{\prime} I^{Q}(t^{\prime})$ is considered instead of the energy per detected particle $Q$, Eq.~(\ref{Qt}) This is due to the quantum-coherent time evolution of the single-particle excitation during its emisision, which gives rise to a negative heat flux at some intermediate times, see Fig.~\ref{ICS}. This non-monotonous behaviour deserves further theoretical investigations for a complete understanding and we refer to a recent work, Ref.~\cite{Ludovico16} that may contribute for additional theoretical insights.

Here, we would like to draw to the attention of the reader that the non-monotonous behavior of the energy per particle may have all its importance when considering coherent single electronic excitations as resources for thermodynamical tasks. Indeed, Figs.~\ref{ELt} and ~\ref{ECSt} show that a single excitation carries a maximum amount of energy at a specific time. By adjusting accordingly the detection process, one could think of extracting an optimal amount of energy from the particle. This is of particular interest with respect to recent works in quantum thermodynamics, aiming at characterizing and extracting energy from quantum coherent processes. \cite{Frenzel:2014,{Korzekwa:2016}}.

The non-monotonous behavior of the energy of the particle also rises the question of the amount of information that is carried by the single electron. It is well-known in classical thermodynamics that energy and information are closely related; for a single particle that can be in two states, the Shannon entropy and the Boltzmann entropy  are proportional up to the Boltzmann constant $k_B$. Our result puts forward the question of the validity of this relation for coherent single excitations. One of the questions of interest could therefore be 
whether 
the information carried by the electron varies in time. 

In the following, we consider the previous questions as motivations for future works and rather concentrate on the quantum properties of the emitted single excitations that can be extracted from transport measurements.

\subsection{Transport tomography of a wave function}

Together, time-dependent charge and heat currents provide complete information on a single-electron wave function in one dimension.
If we represent the wave function in terms of its amplitude $A(t)$ and phase $ \phi(t)$,
\begin{subequations}
\label{wf}
\begin{eqnarray}
 \Psi(t) = A(t) e^{-i \phi(t)}, 
\label{wfa}
\end{eqnarray}
\noindent
then, using Eqs.~(\ref{curra}) and (\ref{currb}) we find,
\begin{eqnarray}
A(t) = \sqrt{\frac{ I(t) }{ e v_{ \mu}}} , 
\nonumber \\
\label{wfb} \\
\phi(t)  = \frac{e }{ \hbar } \int _{}^{t } dt^{\prime} \frac{ I^{Q}(t^{\prime}) }{ I(t^{\prime})  } .
\nonumber 
\end{eqnarray}
\end{subequations}
\ \\ \noindent
This last equation defines a phase of a wave function up to an irrelevant constant.

As we already mentioned, in the case of an adiabatic emission, the measurement of a time-dependent electrical current alone is sufficient to determine  a wave function.\cite{Moskalets:2015kx}  

Note that the measurement of a time-dependent heat current is still challenging. 
However, the recent progress in heat transport measurements on the nanoscale \cite{Lee:2013fc,Jezouin:2013fx} and the efforts made in a time-resolved thermometry down to a single quanta of energy level \cite{Gasparinetti:2015gr} inspire hope that a time-resolved heat measurement is almost within reach of a present-day experiment. 

As it was demonstrated experimentally, long-time measurements can also provide information on the state of a single-particle wave packet. In particular, the current cross-correlation partition noise measurement in the case of colliding wave packets is able to probe the spatial  extension of wave packets. \cite{Bocquillon:2013dp,Dubois:2013dv,Glattli:2016ep}  
When the two wave packets overlap on a wave splitter, the shot noise becomes suppressed due to the Pauli exclusion principle, which forces the two fermions to go to different outputs, thus suppressing the noise.
\cite{Liu:1998gb} 
Loosely speaking, one can say that one fermion plays a role of a non-penetrable wall for the other one. 
This analogy was experimentally realized in Ref.~\cite{Fletcher:2013kt}, where one wave packet was replaced by a dynamical barrier and the shape of the other wave packet was explored.

Below we show that a similar Pauli suppression should be present for a heat partition noise as well. 
However the associated spatial size is different. That is, an electrical shot noise and a heat shot noise are suppressed differently with increasing overlap. 
This characterizes the difference between the  electrical and the  heat currents, now in presence of a quantum-statistical exchange.

\section{Heat partition noise}

\begin{figure}[b]
\centerline{
\includegraphics[width=74mm]{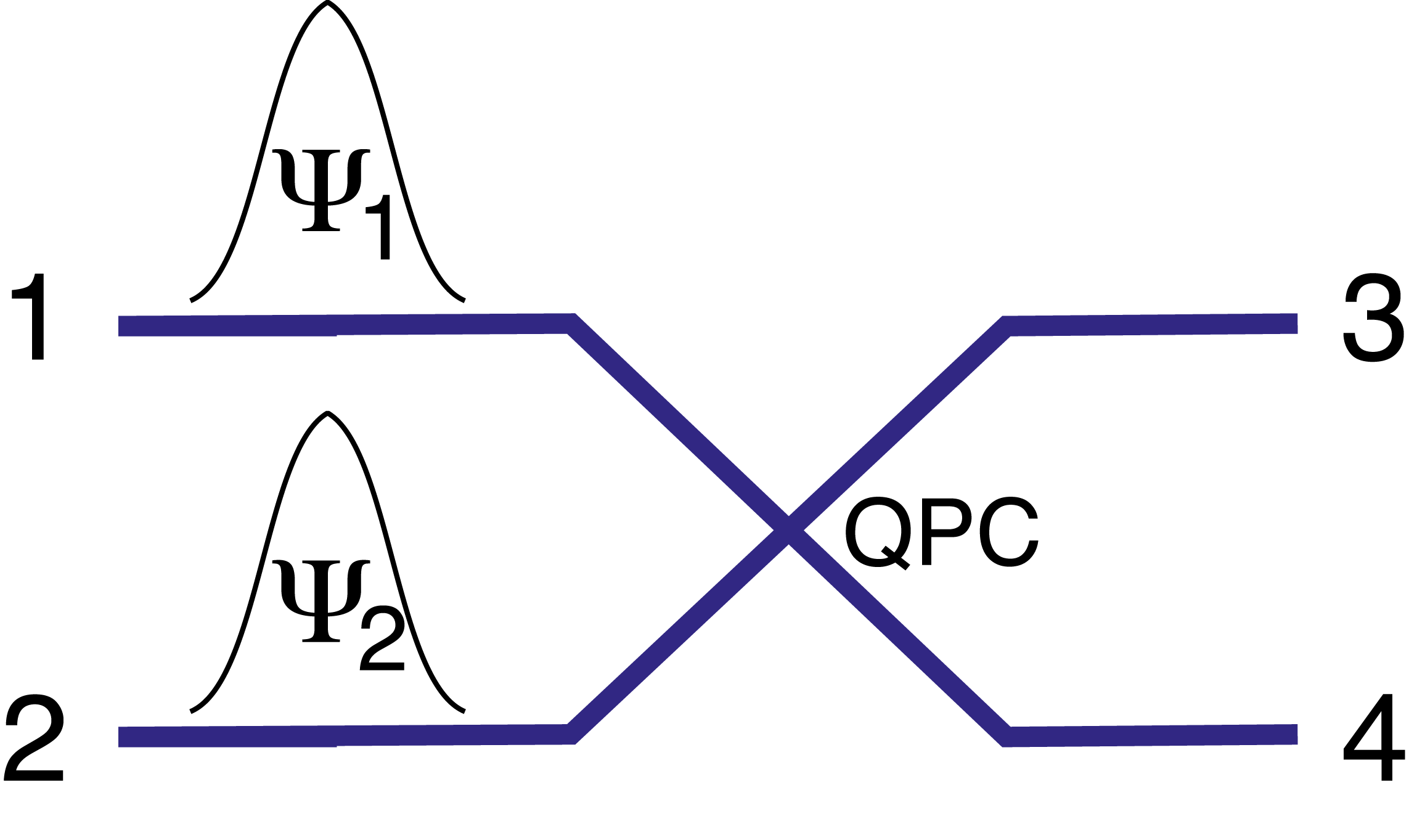}
}
\caption{(Color online) 
A sketch of an electron collider, where two single-particle wave packets $ \Psi_{1}$ and $ \Psi_{2}$ incoming from different input channels $1$ and $2$ can overlap at the quantum point contact QPC before going to either of the two outgoing channels $3$ and $4$. 
 }
\label{QPC}
\end{figure}

Let us consider 
an electronic mesoscopic collider 
\cite{Henny:1999iq,Oliver:1999gb,Oberholzer:2000fy}, where two waveguides come close to each other and electrons from one waveguide can tunnel into the other one through a quantum point contact (QPC), an electronic wave splitter. 
Each incoming channel, $ \alpha = 1,2$, is fed by a single-electron source injecting particles with wave function $ \Psi_{ \alpha}(t) = e^{-i t \frac{ \mu }{ \hbar} } \psi_{ \alpha}(t)$  regularly, at the rate $1/ {\cal T}$, see Fig.~\ref{QPC}. 
Particles emitted during different periods do not overlap with each other.

The transmission $T$ and reflection $R=1-T$ coefficients of the QPC are supposed to be energy-independent.   
Then the time-average correlation function of heat currents 
(see Append.~\ref{hsn}) 
flowing into the two output channels $ \gamma = 3,4$ can be represented as the sum of the three contributions, 

\begin{eqnarray}
{\cal P}^{Q}_{34} = {\cal P}^{Q,1}_{34} + {\cal P}^{Q,2}_{34}+ \delta {\cal P}^{Q}_{34} .
\label{hpn}
\end{eqnarray}
\noindent \\
Here ${\cal P}^{Q, \alpha}_{34}$ is the heat shot noise due to partitioning of electrons emitted by the source $ \alpha$ alone and $\delta {\cal P}^{Q}_{34}$ is the heat correlation noise caused by the collision of particles emitted by different sources at the QPC. 

\subsection{Heat shot noise}

The single-particle heat partition noise is 

\begin{eqnarray}
{\cal P}^{Q,\alpha}_{34} &=& 
- \frac{  RT  }{ {\cal T} }
\left| v_{ \mu} \hbar \int _{- \infty }^{ \infty  } dt \frac{d \psi_{ \alpha}^{*}(t) }{ dt } \psi_{ \alpha}(t)  \right|^{2} 
\nonumber \\
\label{hpn01} \\
&=& -  \frac{ RT  }{ {\cal T} }    Q_{ \alpha} ^{2}, 
\nonumber 
\end{eqnarray} 
\noindent \\
where $Q_{ \alpha} = \int_{- \infty}^{ \infty} dt I_{ \alpha}^{Q}(t) $ is the energy per particle in a long-time limit (when the emission process is completed). 
While calculating the second line in the above equation, we used the identity
$\int _{ }^{  } dt {\rm Re} \left [\psi_{ \alpha}(t)  d \psi_{ \alpha}^{*}(t)/ dt  \right] = 0$ valid at $ \psi_{ \alpha}( - \infty) = \psi_{ \alpha}(+ \infty)$. 
In particular, such an identity holds for a wave function vanishing at long-time limits, $ \psi_{ \alpha}( - \infty) = \psi_{ \alpha}(+ \infty) =0$, of interest here. 

Note that in the case of levitons, the heat shot noise was discussed in Refs.~\cite{Moskalets:2014cr,Battista:2014di}. 

The single-particle heat current cross-correlator can be interpreted in the same way as an electrical current cross-correlator \cite{Blanter:2000kl}. 
Indeed, if we replace $Q_{ \alpha}$ by an electron charge $e$ in Eq.~(\ref{hpn01}), we arrive at the well-known equation for the (electrical) partition noise in the same set-up, ${\cal P}_{34} =- e^{2} RT/ {\cal T}$.\cite{Olkhovskaya:2008en,Bocquillon:2012if,Dubois:2013dv} 
Therefore, what is partitioned are particles, not energy: The heat current cross-correlator describes partitioning of a stream of  indivisible particles carrying each an energy $Q_{ \alpha}$ by a quantum point contact.

\subsection{Heat correlation noise}

The correlation noise is rooted in the quantum-statistical exchange, which is sensitive to the way the two colliding particles will overlap at the QPC. \cite{Buttiker:1990tn,Liu:1998gb,Blanter:2000kl,Splettstoesser:2009im,Moskalets:2011jx,Juergens:2011gu}   
If the overlap is perfect, then the Pauli principle forbids two fermions to go to the same output channel, each particle goes to a separate output. 
As a result, the noise gets suppressed completely. However, if the particles would not overlap, then each of them separately would cause noise.
One can say that the correlation noise suppresses perfectly a single-particle noise (of two particles) such that the total cross-correlation noise vanishes (both electrical and heat). 

The heat correlation noise is 

\begin{subequations}
\label{hpn02}
\begin{eqnarray}
\delta {\cal P}^{Q}_{34} &=& 
2 \frac{ RT   }{ {\cal T} } Q_{1} Q_{2} \left | J^{Q}_{} \right |^{2} ,
\label{hpn02a}
\end{eqnarray}
\ \\ \noindent
where the heat overlap integral is defined as follows,

\begin{eqnarray}
J^{Q}_{} = \frac{ -i \hbar }{ \sqrt{Q_{1} Q_{2}} } v_{ \mu} \int_{- \infty}^{ \infty} dt \frac{d \psi_{1}^{*}(t) }{ dt } \psi_{2}(t) .
\label{hpn02b}
\end{eqnarray}
\end{subequations}
\ \\ \noindent
Note that the squared heat overlap integral preserves symmetry between the sources, $\psi _{1} \leftrightarrow \psi _{2}$. 

If the two sources emit particles, whose wave functions are  the same (hence $Q_{1} = Q_{2}$ and  ${\cal P}^{Q,1}_{34} =  {\cal P}^{Q,2}_{34}$) and whose overlap is perfect at the QPC, then the overlap integral is $J^{Q} = 1$. 
In this case, the heat correlation noise is (minus) twice a single-particle heat shot noise,  $\delta {\cal P}^{Q}_{34} = - 2  {\cal P}^{Q,1}_{34}$, and the total heat noise vanishes, ${\cal P}^{Q}_{34} = 2{\cal P}^{Q,1}_{34} + \delta {\cal P}^{Q}_{34} = 0 $. 

Note that the heat overlap integral $J^{Q}_{}$, which governs the heat noise suppression, 
(see the end of  Append.~\ref{appf})

\begin{eqnarray}
{\cal P}^{Q}_{34} &=& 
-  \frac{ RT   }{ {\cal T} } \left\{ Q_{1}^{2} +  Q_{2}^{2}  - 2 Q_{1} Q_{2} \left | J^{Q}_{} \right |^{2} \right\} ,
\label{hpn-1}
\end{eqnarray}
\ \\ \noindent
is, in general, different from the overlap integral 

\begin{eqnarray}
J_{} = v_{ \mu} \int _{- \infty}^{ \infty } dt \psi_{1}^{*}(t) \psi_{2}(t),
\label{hpn03}
\end{eqnarray}
\ \\ \noindent
which governs an electrical noise suppression in the same setup, \cite{Olkhovskaya:2008en,Jonckheere:2012cu,Dubois:2013fs,Bocquillon:2013fp,Moskalets:2013dl}

\begin{eqnarray}
{\cal P}^{}_{34} = -2 \frac{  RT }{ {\cal T} } e^{2} \left( 1 -  \left | J_{} \right |^{2}\right) ,
\label{hpn04}
\end{eqnarray}
\ \\ \noindent
The difference between the overlap integrals $J^{Q}$ and $J$ is rooted in the difference between the time-resolved currents $I^{Q}(t)$ and $I(t)$.

\subsubsection{Heat versus charge overlap integrals}

As an illustration, let us evaluate the overlap integrals $J^{Q}$, Eq.~(\ref{hpn02}), and $J^{}$, Eq.~(\ref{hpn03}), in the case where the two excitations of the same kind approach the QPC with a time delay $ \tau$. 

For excitations with a wave function $\psi_{Tr}(t)$, Eq.~(\ref{Tr-wf}), we use $ \psi_{1}(t) = \psi_{Tr}(t)$ and $\psi_{2}(t) = \psi_{Tr}(t+\tau)$ and calculate

\begin{eqnarray}
\left | J^{Q}_{Tr} ( \tau) \right |^{2} = \left | J_{Tr} ( \tau) \right |^{2} =e^{- \frac{ | \tau| }{ \tau_{D} }} .
\label{JTr} 
\end{eqnarray}
\ \\ \noindent
In this case, the heat and charge overlap integrals are the same, what is in agreement with the fact that a heat current is proportional to a charge current and they both have the same spatial extension. 

This is no longer the case for colliding levitons. 
Using $ \psi_{1}(t) = \psi_{L}(t)$ and $\psi_{2}(t) = \psi_{L}(t+\tau)$, where $ \psi_{L}(t)$ is given by Eq.~(\ref{L-wf}), we calculate 

\begin{subequations}
\label{JL}
\begin{eqnarray}
\left | J^{Q}_{L} ( \tau) \right |^{2} = \frac{1  }{  \left [ \left( \frac{ \tau }{ 2  \Gamma _{\tau} } \right)^{2} + 1 \right]^{2} } ,
\label{JLa}
\end{eqnarray}

\begin{eqnarray}
\left | J_{L} ( \tau) \right |^{2} = \frac{1  }{   \left( \frac{ \tau }{ 2  \Gamma _{\tau} } \right)^{2} + 1  } .
\label{JLb} 
\end{eqnarray}
\end{subequations}
\ \\ \noindent
The fact that the heat overlap integral as a function of the time delay $ \tau$ is sharper compared to the charge overlap integral is correlated with the fact that a heat pulse $I^{Q}_{L}(t)$ is sharper compared to a charge pulse, $I^{}_{L}(t)$, of a leviton, see Eq.~(\ref{Jlaw}) and the dashed lines in Fig.~\ref{ICS} panels a) and  b).

\begin{figure}[b]
\centerline{
\includegraphics[width=80mm]{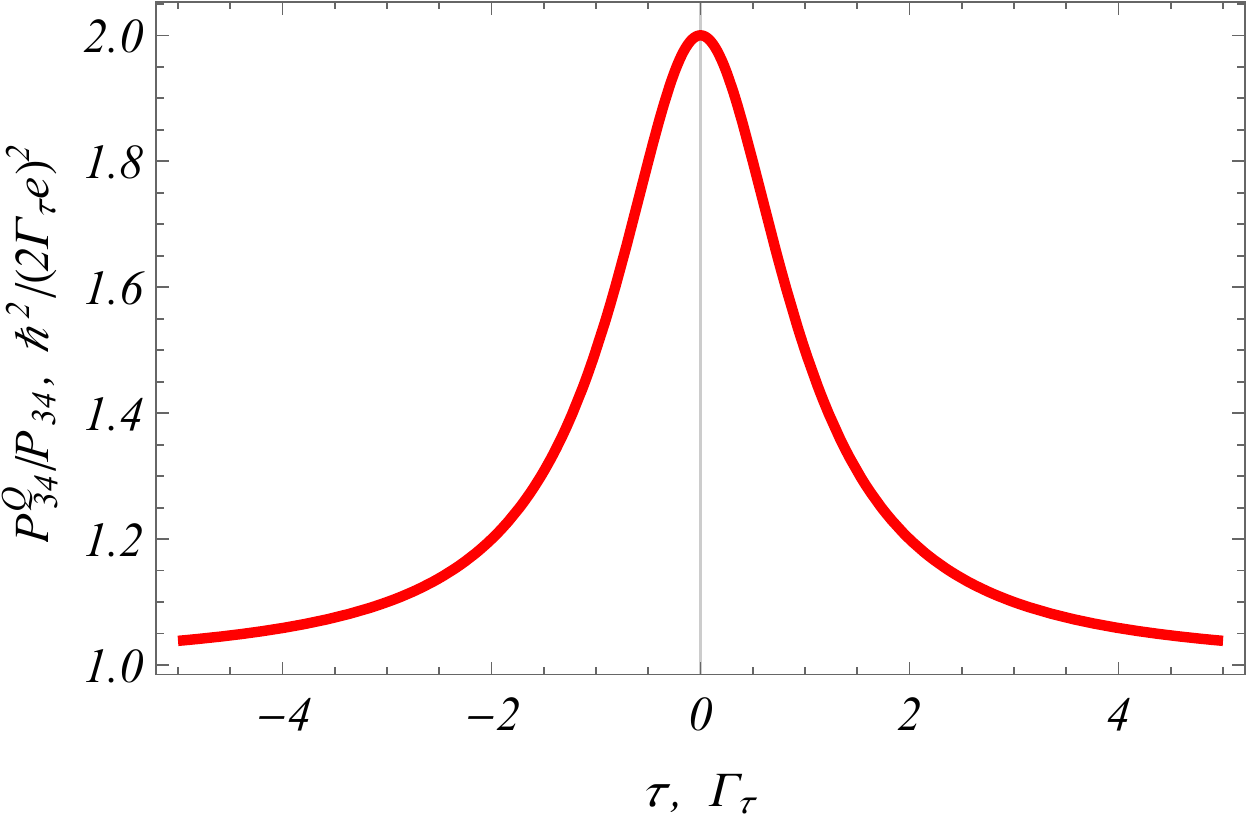}
}
\caption{(Color online) 
The ratio of the heat cross-correlator, ${\cal P}_{34}^{Q}$, Eq.~(\ref{hpn03}), to the electrical cross-correlator, ${\cal P}_{34}$, Eq.~(\ref{hpn04}), for two colliding levitons as a function of a time delay, $ \tau$.  
 }
\label{PQPIL}
\end{figure}

Comparing $\left | J_{} ( \tau) \right |^{2}$ and $\left | J_{}^{Q} ( \tau) \right |^{2}$, we see that there is an interval of time delays $\tau$ where the charge partitioning is  suppressed while the heat partitioning is not. 
As a consequence, the heat partition noise becomes enhanced  relatively to a charge partition noise, see Fig.~\ref{PQPIL}.
That is, if the particles overlap only partially, then the energy carried by scattered particles fluctuates more than the charge carried by them. 
These additional fluctuations could be attributed to the fact that, in the case of heat, the quantum-statistical exchange is not constrained by the conservation law per particle as in the case of charge. 

We see that the difference in nature between heat and charge single-particle currents can be observed not only in a time-resolved measurement in the near field region of the source but also in a long-time measurement as soon as the  quantum-statistical exchange is involved. 

For the case of an electron emission from a level moving 
at a constant rapidity, 
see Eq.~(\ref{CS-wf}), the overlap integrals  are given by the same equations as for a leviton, $J_{CS}^{Q} = J_{L}^{Q}$ and $J_{CS} = J_{L}$.  
This means, in particular, that the overlap integral (hence a collision experiment) provides only a partial information on the state, even on the density profile.

\section{Conclusion}

We discussed a relation between the quantum-mechanical characteristics of single-particle excitations injected into  
a mesoscopic ballistic conductor 
and the transport measurements which can be performed at the output 
of such a conductor. 

Using the Floquet scattering matrix approach to quantum transport, 
we expressed the time-dependent electrical and heat currents flowing out of a multi-terminal 
conductor,
as well as their correlation functions averaged over time, in terms of the excess electronic correlation function, which describes either single- or multi-particle excitations injected by an electronic source on the top of the Fermi sea of non-interacting electrons at a zero or a finite temperature. 

As an illustration we considered  single-electron injections and analyzed respective transport characteristics. 
We found that unlike the electrical current, the heat current associated to  specific single-electron states can attain a negative value at short times, see Fig.~(\ref{ICS}), 
panel b), a blue line. 
Here negative means that an energy seems to flow into an electron source instead of to flow out of it. 
This counterintuitive fact suggests that a heat current is not directly measurable. 
We argue that this is really the case. 
Indeed, to measure a heat current, we need to perform two successive measurements of energy at close times on the same state. 
However, the first measurement affects the single-particle state such that the second measurement is meaningless. 
As a result, the measurable 
quantity, 
mostly discussed in the main text, 
is a single-particle energy rather then an energy flux. 
This interpretation is in line with the interpretation of a single-particle electrical current, whose dependence with time provides us with a probability density to register the total charge of an electron rather than with a charge density distribution.  

However there is also a difference between a charge and an energy detection. 
A charge measurement gives always a fixed value, an electron charge.
In contrast, an energy measurement depends essentially on the distance between the source and the detector. 
If the distance is larger than a size of an emitted wave packet, then the energy detected is fixed. 
In contrast, if such a distance is smaller than a size of an emitted wave packet, then the energy measured by a detector depends on an actual time, \textit{i.e.} the time at which a particle is detected. 
This time-dependent energy is given by the heat current, integrated  from a time of emission up to a time of detection [and properly normalized, see Eq.~(\ref{Qt})]. 
The use of such a time-dependent energy measured on the ensemble of identically prepared single-particle states allows us to calculate a time-dependent heat current. 
In view of this interpretation, a negative heat current just means that an energy of a single-particle measured at earlier times can exceed an energy measured at later times, see Figs.~\ref{ELt} and \ref{ECSt}. 

As we demonstrated, the knowledge of a time-dependent electrical current and a time-dependent heat current permits the full reconstruction of a single-particle wave function, see Eqs.~(\ref{wf}). 

The above mentioned difference between an electrical current and a heat current can also be brought to light using a current cross-correlation function averaged over time, that does not require a challenging time-resolved measurement.  
This difference appears if the two identical wave packets collide at a wave splitter. 
The Pauli principle forbids two identical fermions to be at the same place. 
Therefore, two perfectly overlapping electrons are necessarily scattered to two different outputs. 
As a result the partition noise gets suppressed and a cross-correlation function for outgoing current nullifies. 
However, if the overlap is not perfect, the partition noise is suppressed only partially since sometimes two electrons can be scattered into the same output channel while nothing is scattered to the other one. 
Surprisingly, the partial suppression of an electrical noise and a heat noise are governed by two different overlap integrals, whose dependence on wave functions of colliding particles reflects the difference between the  corresponding time-dependent currents, compare Eqs.~(\ref{hpn02b}) and (\ref{currb}) as well as Eqs.~(\ref{hpn03}) and (\ref{curra}).

\begin{acknowledgement}
We thank David S\'{a}nchez and Liliana Arrachea for useful comments on the manuscript. 
G.H. acknowledges support from the SNF through the NCCR QSIT and through the Marie Heim-V\"ogtlin grant no. 18874.
\end{acknowledgement}

\appendix

\section{Excess first-order correlation matrix}
\label{appa}

In this appendix, we use the Floquet scattering matrix theory \cite{Moskalets:2011cw} to express currents flowing through a mesoscopic system in terms of quantum-mechanical characteristics of electron excitations injected into such a system. 
The presented theory is valid for either single- or multi-particle pure or mixed incoming states.  

The system we consider is an electronic multi-terminal 
ballistic conductor 
made of  single-channel chiral waveguides \cite{Buttiker:2009bg} for non-interacting and spinless electrons originating from metallic contacts and of quantum point contacts where the two waveguides come close to each other\cite{Beenakker:1991uq}, playing the role of wave splitters. 
The 
conductor 
is connected to $N_{r}$ metallic contacts, electronic reservoirs. 
An electron system at each contact $ \alpha = 1, \dots, N_{r}$ is at equilibrium,   described by the Fermi distribution function $f_{ \alpha}$ with the chemical potential $ \mu_{ \alpha}$ and the temperature $ \theta_{ \alpha}$. 

Electrons are injected into the 
conductor 
by a periodically working source, which  emits a stream of particles.
We characterize a source connected to the lead $ \alpha$  by the Floquet scattering matrix $S_{F,\alpha}$. 
Examples of such sources are (i) a quantum dot side-attached to an incoming waveguide \cite{Feve:2007jx} or (ii) a time-dependent voltage pulse applied directly to the reservoir, from which the waveguide comes from \cite{Dubois:2013dv}.  
In general, 
the conductor can be fed by  
one or several electronic sources, which are all driven by potentials having the same period ${\cal T}$.

The state of the particles injected by the source into a \textit{single} chiral waveguide is conveniently characterized by the excess electronic correlation function.\cite{Grenier:2011js,Grenier:2011dv,Haack:2013ch} 
Here, we introduce the excess electronic correlation \textit{matrix} to describe the state of particles leaving a multi-terminal 
conductor. 

The first-order correlation matrix ${\pmb {\cal  G}}_{out}^{(1)}\left(1;2  \right)$ for electrons leaving the multi-terminal 
conductor 
has elements ${\cal G}^{(1)}_{ \alpha \beta}\left(1;2  \right)$  defined as follows, 

\begin{eqnarray}
{\cal G}^{(1)}_{ \alpha \beta}\left(1;2  \right) =
\langle \hat\Psi^{\dag}_{ \alpha}(1)  \hat\Psi_{ \beta}(2) \rangle ,  
\label{CM-01}
\end{eqnarray}
\ \\ \noindent
where $\hat\Psi_{ \alpha} (j) \equiv \hat\Psi_{ \alpha}\left(x_{j}t_{j} \right)$ is a single-particle electron field operator in second quantization evaluated at point $x_{j}$ and time  $t_{j}$,  $j=1,2$ in the outgoing waveguide $ \alpha$ after the 
conductor. 
The quantum statistical average $\langle \dots \rangle$ is made over the equilibrium state of electrons incoming from the metallic contacts. 
The correlation matrix  ${\pmb {\cal  G}}_{out}^{(1)}$ contains information about both electrons of the Fermi sea and particles injected by the sources. 
To access the information that concerns solely the particles emitted by the sources, we introduce {\it the excess first-order correlation matrix}, which is evaluated as the difference of the electronic correlation matrices with the sources \textit{switched on} and \textit{off},

\begin{eqnarray}
{\pmb G}_{out}^{(1)}\left(1;2  \right) = {\pmb {\cal  G}}_{out,on}^{(1)}\left(1;2  \right) - {\pmb {\cal  G}}_{out,off}^{(1)}\left(1;2  \right) .
\label{CM-02}
\end{eqnarray}

\subsection{Excess correlation matrix in terms of the Floquet scattering matrix}

The next step is to express the elements of the correlation matrix in terms of the Floquet scattering matrix characterizing the electronic sources.  
To this end we first introduce the field operator in second quantization  $\hat\Psi\left(x_{j}t_{j} \right)$ for electrons in an electrical conductor \cite{Buttiker:1992ge}. 
For chiral electrons in lead $ \alpha$ it reads 

\begin{eqnarray}
\hat\Psi_{ \alpha}\left(x_{j}t_{j} \right) = \int \frac{dE }{\sqrt{h v_{ \alpha}(E)} } e^{i\phi_{j, \alpha}(E) } \hat b_{ \alpha}\left( E \right) .
\label{CM-03}
\end{eqnarray}
\ \\ \noindent
Here $1/[h v_{ \alpha}(E)]$ is a one-dimensional density of states at energy $E$,  $\hat b_{ \alpha}(E)$ is an annihilation operator for electrons leaving the 
conductor 
through the waveguide $\alpha$, and the phase $\phi_{j, \alpha}(E) = -E t_{j}/\hbar + k_{ \alpha}(E) x_{j}$.   
Then we use the stationary scattering matrix of a multi-terminal 
ballistic quantum conductor 
$ {\bf S}_{C}(E)$ and the Floquet scattering matrices of the sources, ${\bf S}_{F, \gamma}$, and relate the $\hat b_{ \alpha}$-operators to the $\hat a_{ \gamma}$-operators which describe equilibrium electrons coming from the metallic contacts \cite{Moskalets:2002hu},

\begin{eqnarray}
\hat b_{ \alpha}(E) = \sum_{ \gamma=1}^{N_{r}}
\sum_{n=-\infty}^{\infty} S_{C, \alpha \gamma}\left( E \right) S_{F, \gamma}\left( E, E_{n} \right) \hat a_{ \gamma}\left( E_{n} \right) . 
\label{CM-04}
\end{eqnarray}
\ \\ \noindent 
Here $E_{n} = E + n \hbar \Omega$, where $ \Omega = 2 \pi/{\cal T}$.
The Floquet scattering matrix element $S_{F, \gamma}\left( E, E_{n} \right)$  is a quantum mechanical amplitude for an electron with energy $E_{n}$ in an incoming waveguide $ \gamma$ to emit (or absorb) $n>0$ (or $n<0$) energy quanta $\hbar\Omega$ while passing by the source. 
If the source is off or if there is no an electronic source in the lead $ \gamma$, then ${\bf S}_{F, \gamma}$ becomes the unit matrix with elements $S_{F, \gamma}(E,E_{n}) =  \delta_{n0}$, where $\delta_{n0}$ is the Kronecker symbol ($1$ for $n=0$ and $0$ otherwise).  

In order to perform the quantum-statistical average in Eq.~(\ref{CM-01}), we use the following relation $\left\langle \hat a^{\dag}_{ \gamma}(E) \hat a_{ \gamma^{\prime}}(E^{\prime}) \right\rangle = f_{ \gamma}(E) \delta_{ \gamma \gamma^{\prime}} \delta\left( E - E^{\prime} \right)$, which is valid since electrons of a metallic contact are at equilibrium.
Here $\delta\left( E - E^{\prime} \right)$ is the Dirac delta-function.
Finally, using the quantities introduced above, we represent the elements of the excess correlation matrix ${\bf G}^{(1)}_{out}$ in terms of the Floquet scattering matrices of the sources, ${\bf S}_{F, \gamma}$, $ \gamma = 1, \dots N_{r}$,  and the scattering matrix of the 
conductor, 
${\bf S}_{C}$, 

\begin{eqnarray}
G^{(1)}_{out, \alpha \beta}(1;2) = 
\sum_{ \gamma = 1 }^{ N_{r} } 
\sum_{n,m}^{} 
\int \frac{dE f_{ \gamma}\left( E \right) e^{-i\phi_{1, \alpha}(E_{n})} e^{i\phi_{2, \beta}(E_{m})}   }{ h \sqrt{v_{ \alpha}\left( E_{n}  \right) v_{ \beta}\left(  E_{m}  \right) } } 
\nonumber \\
\label{CM-05} \\
\times  
\Big\{ 
S_{C, \alpha \gamma}^{*}\left(E_{n}\right)
S_{F, \gamma}^{*}\left(E_{n},E  \right)
S_{C, \beta \gamma}\left(E_{m} \right)
S_{F, \gamma}\left(E_{m},E  \right)
\nonumber \\
- 
\delta_{ \alpha \beta}  
\delta_{m,0}
\delta_{n,0}  
\left | S_{C, \alpha \gamma} \right |^{2}
 \Big\} .
\nonumber 
\end{eqnarray}
\ \\ \noindent
The matrix with the above elements  satisfies the symmetry 

\begin{eqnarray}
{\bf G}^{(1)}_{out}( t_{1};t_{2})  = \left [ {\bf G}^{(1)}_{out}( t_{2};t_{1})  \right]^{ \dag}.
\label{sym-1}
\end{eqnarray}
\ \\ \noindent

In the following, we are interested in the case where all the contacts $ \gamma$ are characterized by the same chemical potential and the same  temperature, $ \mu_{ \gamma} = \mu$,  $\theta_{ \gamma} = \theta$, $ f_{ \gamma}(E) = f_{0}(E)$,  $\forall \gamma$.  
Let us mention that it is possible to incorporate additional DC or AC potentials  $V_{ \gamma}(t)$ at any contact $ \gamma$  by adding a phase factor $\exp\left\{ - ie/\hbar \int _{}^{t } dt^{\prime} V_{ \gamma}(t^{\prime}) \right\}$ to the corresponding scattering matrix of an electronic source ${\bf S}_{F, \gamma}$.

\subsection{A linear dispersion approximation}

We suppose that all the relevant energy scales (such as applied voltage biases, energies of emitted electrons, energy quanta $ \hbar \Omega$,  temperature, etc.) of the problem are much smaller than the chemical potential $ \mu$. 
Then, for energies close to the Fermi energy we can linearize the dispersion relation, 

\begin{eqnarray}
k_{ }\left( E_{n} \right) \approx k_{ \mu}  + \frac{ \epsilon + \hbar n\Omega }{ \hbar v_{ \mu} } ,
\label{CM-07}
\end{eqnarray}
\ \\ \noindent
and represent the corresponding phases as follows,

\begin{eqnarray}
\phi_{j}(E_{n}) =   \phi_{ j,\mu } -  \left( t_{j} - \frac{x_{j} }{ v_{ \mu} } \right) \frac{ \epsilon + \hbar n \Omega }{ \hbar } .
\label{CM-08}
\end{eqnarray}
\ \\ \noindent
Here $ \phi_{ j, \mu} = - \mu t_{j}/\hbar + k_{ \mu} x_{j}$ and $k_{ \mu}$ are respectively the  phase factor and the wave vector for electrons with Fermi energy $ \mu$; $ \epsilon = E - \mu$ is an energy counted from the Fermi energy. 
For shortness,  we denote below the difference  $ t_{j} - x_{j}/v_{ \mu}$ simply as $t_{j}$. 
Note that we consider the dispersion relation to be the same in all leads, hence we drop the index $ \alpha$.

Within a linear dispersion approximation, the elements of the correlation matrix become 

\begin{eqnarray}
G^{(1)}_{ out,\alpha \beta}( t_{1};t_{2}) = 
\frac{ 1 }{hv_{ \mu} }
\int   dE f_{0}(E)
\sum_{n,m}^{} 
e^{i t_{1}  \frac{ E_{n} }{\hbar} } 
e^{ - i t_{2}  \frac{ E_{m} }{\hbar} } 
\nonumber \\
\label{CM-06} \\
\times  
\Bigg\{ 
\sum_{ \gamma = 1 }^{ N_{r} } 
S_{C, \alpha \gamma}^{*}\left(E_{n}\right)
S_{F, \gamma}^{*}\left(E_{n},E  \right)
\nonumber \\
\times
S_{C, \beta \gamma}\left(E_{m} \right)
S_{F, \gamma}\left(E_{m},E  \right)
- 
\delta_{ \alpha \beta}  \delta_{m,0}\delta_{n,0} 
\Bigg\} 
\nonumber \\
= 
\frac{ 1 }{hv_{ \mu} } \int   dE 
\sum_{n,m}^{} 
e^{i t_{1}  \frac{ E_{n} }{\hbar} } 
e^{ - i t_{2}  \frac{ E_{m} }{\hbar} } 
\left\{ f_{0}(E) - f_{0}(E_{n}) \right\}
\nonumber \\
\sum_{ \gamma = 1 }^{ N_{r} } 
S_{C, \alpha \gamma}^{*}\left(E_{n}\right)
S_{F, \gamma}^{*}\left(E_{n},E  \right)
S_{C, \beta \gamma}\left(E_{m} \right)
S_{F, \gamma}\left(E_{m},E  \right)
.
\nonumber 
\end{eqnarray}
\ \\ \noindent
Here, we additionally took into account the unitarity of the scattering matrix describing the 
conductor, 
${\bf S}_{C}^{\dag}{\bf S}_{C} = {\bf 1} \Rightarrow \sum_{ \gamma = 1 }^{N_{r}} \left | S_{C, \alpha \gamma} \right |^{2} = 1$.

\section{Electrical current in terms of the excess correlation matrix}

\label{eFl}

The second-quantization operator of an electrical current flowing in lead $ \alpha$ reads, \cite{Buttiker:1992ge}

\begin{eqnarray}
\hat{I}_{\alpha}(t) & = & \frac{e}{h} \iint  dE dE^\prime e^{i \frac{E - E^\prime}{\hbar} t} \nonumber \\
\label{e-01} \\
&& \times \left\{ b_{\alpha}^\dag(E) b_{\alpha}(E^\prime) - a_{\alpha}^\dag(E) a_{\alpha}(E^\prime) \right\} . \nonumber
\end{eqnarray}
\ \\ \noindent
We use Eq.~(\ref{CM-04}) and find a time-dependent electrical current $I_{ \alpha}(t) = \left\langle \hat{I}_{\alpha}(t) \right\rangle$, \cite{Moskalets:2007dl}

\begin{eqnarray}
I_{ \alpha}(t) = 
\frac{ e }{h  }
\int   dE f_{0}(E)
\sum_{n,m}^{} 
e^{i t  \frac{ E_{n} - E_{m} }{\hbar} } 
\Bigg\{ 
- \delta_{m,0}\delta_{n,0} 
+ \sum_{ \gamma = 1 }^{ N_{r} } 
\nonumber \\
\label{e-02} \\
S_{C, \alpha \gamma}^{*}\left(E_{n}\right)
S_{F, \gamma}^{*}\left(E_{n},E  \right)
S_{C, \alpha \gamma}\left(E_{m} \right)
S_{F, \gamma}\left(E_{m},E  \right)
 \Bigg\} 
 \nonumber \\
= \frac{ e }{h  }\int   dE 
\sum_{n,m}^{} 
e^{i t  \frac{ E_{n} - E_{m} }{\hbar} } 
\left\{ f_{0}(E) - f_{0}(E_{n}) \right\} 
\sum_{ \gamma = 1 }^{ N_{r} } 
\nonumber \\
S_{C, \alpha \gamma}^{*}\left(E_{n}\right)
S_{F, \gamma}^{*}\left(E_{n},E  \right)
S_{C, \alpha \gamma}\left(E_{m} \right)
S_{F, \gamma}\left(E_{m},E  \right)
.
\nonumber 
\end{eqnarray}
\ \\ \noindent
Comparing the above equation and Eq.~(\ref{CM-06}), we find

\begin{eqnarray}
I_{ \alpha}(t) = e v_{ \mu} G^{(1)}_{ out,\alpha \alpha}( t; t) . 
\label{e-03}
\end{eqnarray}
\ \\ \noindent
This equation is a generalization of the result derived in Ref.~\cite{Moskalets:2016dx}, valid for the single-channel case.

\section{Heat current in terms of the excess correlation matrix}

\label{hFl}

The second-quantization operator of a heat current flowing in lead $ \alpha$ reads, \cite{Battista:2013ew,{Lim:2013de},{Ludovico:2014de},{Moskalets:2014cr},{Battista:2014di}}

\begin{eqnarray}
\hat{I}_{\alpha}^{Q}(t) & = & 
\frac{1}{h} \iint  dE dE^\prime 
\left( \dfrac{E + E^\prime}{2} - \mu \right)
e^{i \frac{E - E^\prime}{\hbar} t} \nonumber \\
\label{h-01} \\
&& \times \left\{ b_{\alpha}^\dag(E) b_{\alpha}(E^\prime) - a_{\alpha}^\dag(E) a_{\alpha}(E^\prime) \right\} . \nonumber
\end{eqnarray}
\ \\ \noindent
A heat current is defined in the standard manner as an energy current minus the chemical potential $ \mu$, multiplied by a particle current.  
We use Eq.~(\ref{CM-04}) and find a time-dependent heat current $I_{ \alpha}^{Q}(t) = \left\langle \hat{I}_{\alpha}^{Q}(t) \right\rangle$, 

\begin{eqnarray}
I_{ \alpha}^{Q}(t) = 
\frac{ 1 }{h  }
\int   dE f_{0}(E)
\sum_{n,m}^{} 
e^{i t  \frac{ E_{n} - E_{m} }{\hbar} } 
\left( \dfrac{E_{n} + E_{m} }{2} - \mu \right) 
\nonumber \\
\label{h-02} \\
\times
\Bigg\{ 
\sum_{ \gamma = 1 }^{ N_{r} } 
S_{C, \alpha \gamma}^{*}\left(E_{n}\right)
S_{F, \gamma}^{*}\left(E_{n},E  \right)
\nonumber \\
S_{C, \alpha \gamma}\left(E_{m} \right)
S_{F, \gamma}\left(E_{m},E  \right)
- \delta_{m,0}\delta_{n,0} 
 \Bigg\} .
\nonumber 
\end{eqnarray}
\ \\ \noindent
Note that in the adiabatic regime a time-dependent heat current was discussed in Refs.~\cite{{Avron:2001fl},{Avron:2004vb}}. 

Comparing this equation and Eq.~(\ref{CM-06}), we find the general expression of the heat current:

\begin{equation}
I_{ \alpha}^{Q}(t) =
v_{ \mu} \left\{ \frac{ -i \hbar }{ 2} \left( \frac{ \partial }{ \partial t } - \frac{ \partial }{ \partial t^{\prime} } \right) -  \mu \right\} G^{(1)}_{out, \alpha \alpha}(t;t^{\prime}) \bigg|_{t = t^{\prime}}. 
\label{h-03} 
\end{equation}

Let us remark that the excess correlation function $G^{(1)}_{out, \alpha \alpha}$ is represented in terms of electronic wave functions when the outgoing state is a pure state.\cite{{Grenier:2013gg},{Moskalets:2015kx}}
In particular, just behind the source emitting a single particle in a pure state with wave function $ \Psi(t)$, the correlation function reads

\begin{eqnarray}
G^{(1)}_{ }( t_{1}; t_{2}) = \Psi_{ }^{*}(t_{1}) \Psi_{ }(t_{2}). 
\label{h-04}
\end{eqnarray}
\noindent \\
Equations (\ref{e-03}) and (\ref{h-03}) then lead to Eqs.~(\ref{curr}).

\section{Electrical noise in terms of the excess correlation matrix}

\label{appd}

The correlation function of electrical currents flowing into leads $ \alpha$ and $ \beta$, averaged over two times, is defined as follows,
\cite{Blanter:2000kl}

\begin{eqnarray}
{\cal P}_{ \alpha \beta} &=& 
\frac{1 }{2 } 
\int _{0}^{ {\cal T} } \frac{dt }{ {\cal T} } \int _{- \infty}^{ \infty } d \tau 
\nonumber \\
\label{en-01} \\
&&
\left\langle \Delta \hat I_{ \alpha}(t) \Delta \hat I_{ \beta}(t + \tau)  + \Delta \hat I_{ \beta}(t + \tau)  \Delta \hat I_{ \alpha}(t)  \right\rangle ,
\nonumber 
\end{eqnarray}
\ \\ \noindent
where $\Delta \hat I_{ \alpha}(t) =  \hat I_{ \alpha}(t) - \left\langle  \hat I_{ \alpha}(t) \right\rangle$ is an operator of current fluctuations. 

The above quantity  is usually referred to as an electrical noise at  zero frequency. 
It consists of  thermal noise and  shot noise.
The former,  present at finite temperatures, is due to fluctuations of the  occupation probability of electrons in the reservoirs.
The latter, also called the partition noise, is due to the granular nature of the  electrical charge. 
The shot noise appears if particles incoming to 
the multi-terminal conductor 
from a single input are scattered to several outputs.  
Here, we are interested in the partition noise and, therefore, consider a zero temperature limit $ \theta = 0$ where the thermal noise vanishes. 

We use Eq.~(\ref{e-01}) in Eq.~(\ref{en-01}) and find at zero temperature, \cite{Moskalets:2004ct}

\begin{eqnarray}
{\cal P}_{\alpha\beta}  =  \dfrac{e^2}{2h} \int dE
\sum\limits_{q=-\infty}^\infty  
\left\{ f_{0}\left(E_{} \right) - f_{0}\left(E_{q} \right) \right\}^{2}   
\nonumber \\
\sum\limits_{\gamma, \delta  = 1}^{N_r } 
\sum\limits_{n,m=-\infty}^\infty  
{\rm S} _{F,\alpha \gamma }^* \left( E _{n}, E \right)\, 
{\rm S} _{F,\beta \gamma }^{} \left( E_{m}, E  \right)\,
\label{en-02} \\
{\rm S} _{F,\beta \delta }^* \left( E_{m} , E_{q}   \right) 
{\rm S} _{F,\alpha \delta }^{} \left( E _{n} , E_{q}   \right)\, 
. 
\nonumber
\end{eqnarray}
\ \\ \noindent
Here, for shortness, we introduce the Floquet scattering matrix that describes the 
conductor 
together with electronic sources, $S_{F, \alpha \gamma}(E_{n}, E) =S_{C, \alpha \gamma}\left( E_{n} \right) S_{F, \gamma}\left( E_{n}, E \right)$.
We remind that  all reservoirs connected to the incoming channels are characterized by the same Fermi distribution function, $f_{ \gamma}(E) = f_{0}(E)$, $ \forall \gamma$.

\subsection{Conservation law for electrical noise}

We use the unitarity property of the Floquet scattering matrix, \cite{Moskalets:2002hu}

\begin{eqnarray}
\sum\limits_{ \beta  = 1}^{N_r } 
\sum\limits_{m=-\infty}^\infty  
{\rm S} _{F,\beta \gamma }^{} \left( E_{m}, E  \right)\,
{\rm S} _{F,\beta \delta }^* \left( E_{m} , E_{q}   \right) 
= \delta_{q0 } \delta_{ \gamma \delta}\, ,
\nonumber \\
\label{en-03} \\
\sum\limits_{ \beta  = 1}^{N_r } 
\sum\limits_{m=-\infty}^\infty  
{\rm S} _{F, \gamma \beta }^{} \left( E_{ }, E_{m}  \right)\,
{\rm S} _{F, \delta \beta }^* \left( E_{q} , E_{m}   \right) 
= \delta_{q0 } \delta_{ \gamma \delta}\, ,
\nonumber 
\end{eqnarray}
\ \\ \noindent
and show that a zero-frequency electrical noise is subject to the following conservation law, \cite{Moskalets:2004ct} 

\begin{equation}
\label{en-04} 
\sum\limits_{ \alpha=1}^{N_{r}} {\cal P}_{\alpha\beta}^{} = 
\sum\limits_{ \beta=1}^{N_{r}} {\cal P}_{\alpha\beta}^{} = 0 .
\end{equation}
\ \\ \noindent
This property is a consequence of the charge conservation.  

Using Eq.~(\ref{en-04}), one can express the auto-correlator in terms of cross-correlators,

\begin{eqnarray}
{\cal P}_{\alpha \alpha}^{} = - \sum\limits_{ \beta \ne \alpha= 1}^{N_{r}} {\cal P}_{\alpha\beta}^{}.
\label{en-05}
\end{eqnarray}
\ \\ \noindent
Below we concentrate on a cross-correlator.

\subsection{Electrical current cross-correlator}

Using Eq.~(\ref{en-03}), we rewrite Eq.~(\ref{en-02}) for $ \alpha \ne \beta$ as follows,

\begin{eqnarray}
{\cal P}_{\alpha \ne \beta}^{}  =  -\dfrac{e^2}{h} 
\sum\limits_{n,m=-\infty}^\infty  
\nonumber \\
\int dE f_{0}\left(E_{} \right)  
\sum\limits_{\gamma = 1}^{N_r } 
{\rm S} _{F,\alpha \gamma }^* \left( E _{n}, E \right)\, 
{\rm S} _{F,\beta \gamma }^{} \left( E_{m}, E  \right)\,
\label{en-06} \\
\sum\limits_{q=-\infty}^\infty   f_{0}\left(E_{q} \right) 
\sum\limits_{\delta  = 1}^{N_r } 
{\rm S} _{F,\alpha \delta }^{} \left( E _{n} , E_{q}   \right)
{\rm S} _{F,\beta \delta }^* \left( E_{m} , E_{q}   \right) 
\, . 
\nonumber
\end{eqnarray}

\subsubsection{A long-period driving limit}
\label{ldpl}

To proceed, we first assume the period ${\cal T}$ of the drive to be long enough, such that particles emitted during different periods do not overlap.  Hence, they are uncorrelated. 
Mathematically, this implies that one can go from a discrete variable $q$ which defines an energy $E_{q} = E + q \hbar \Omega$, to a continuous energy variable denoted $E_{q}$. 
Such a change of variables is realized by the following substitutions. 

\begin{eqnarray}
\sum\limits_{q=-\infty}^{\infty} \to \int  \frac{ d E_{q} }{ \hbar \Omega } , \quad
\int _{0}^{ {\cal T} } dt \to \int _{- \infty}^{ \infty } dt ,
\label{en-07}
\end{eqnarray}
\ \\ \noindent 
Then Eq.~(\ref{en-06}) becomes, 

\begin{eqnarray}
{\cal P}_{\alpha \ne \beta}^{}  =  
-\dfrac{e^2}{h^{2} {\cal T}} 
\sum\limits_{n,m=-\infty}^\infty  
\nonumber \\
\int dE f_{0}\left(E_{} \right)  
\sum\limits_{\gamma = 1}^{N_r } 
{\rm S} _{F,\alpha \gamma }^* \left( E _{n}, E \right)\, 
{\rm S} _{F,\beta \gamma }^{} \left( E_{m}, E  \right)\,
\label{en-08} \\
\nonumber \\
\int _{}^{ } dE_{q} f_{0}\left(E_{q} \right) 
\sum\limits_{\delta  = 1}^{N_r } 
{\rm S} _{F,\alpha \delta }^{} \left( E _{n} , E_{q}   \right)
{\rm S} _{F,\beta \delta }^* \left( E_{m} , E_{q}   \right) 
\, . 
\nonumber
\end{eqnarray}
\ \\ \noindent
Comparing this expression and Eq.~(\ref{CM-06}) for $ \alpha \ne \beta$, we arrive at the relation we are looking for, namely

\begin{eqnarray}
{\cal P}_{\alpha \ne \beta}^{}  &=& 
- \frac{ e^{2} v_{ \mu}^{2} }{ {\cal T} }
\int _{0}^{ {\cal T} } dt \int _{- \infty}^{ \infty } d \tau 
\nonumber \\
&&
G^{(1)}_{ out,\alpha  \beta}( t;t+ \tau)  
G^{(1)}_{ out, \beta \alpha}(t+ \tau; t)   
\label{en-09} \\
&=& - \frac{ e^{2} v_{ \mu}^{2} }{ {\cal T} }
\int _{0}^{ {\cal T} } dt \int _{- \infty}^{ \infty } d \tau 
\left | G^{(1)}_{ out,\alpha  \beta}( t;t+ \tau)  \right |^{2} .
\nonumber 
\end{eqnarray}
\ \\ \noindent
The electrical cross-correlator is clearly negative, as it should be.\cite{Buttiker:1992ge}
We  remind that  this equation implies the limit ${\cal T} \to \infty$. 

The auto-correlation noise, ${\cal P}_{ \alpha \alpha}$, is expressed in terms of all cross-correlation contributions according to Eq.~(\ref{en-05}).

\section{Heat noise in terms of the excess-correlation matrix}

\label{hsn}

By analogy with an electrical current correlation function, Eq.~(\ref{en-01}), we define the correlation function of heat currents averaged over time,

\begin{eqnarray}
{\cal P}_{ \alpha \beta}^{Q} &=& 
\frac{1 }{2 } 
\int _{0}^{ {\cal T} } \frac{dt }{ {\cal T} } \int _{- \infty}^{ \infty } d \tau 
\nonumber \\
\label{hn-01} \\
&&
\left\langle \Delta \hat I_{ \alpha}^{Q}(t) \Delta \hat I_{ \beta}^{Q}(t + \tau)  + \Delta \hat I_{ \beta}^{Q}(t + \tau)  \Delta \hat I_{ \alpha}^{Q}(t)  \right\rangle ,
\nonumber 
\end{eqnarray}
\ \\ \noindent
where $\Delta \hat I_{ \alpha}^{Q}(t) =  \hat I_{ \alpha}^{Q}(t) - \left\langle  \hat I_{ \alpha}^{Q}(t) \right\rangle$ is an operator of heat current fluctuations. 
The heat current operator $\hat I_{ \alpha}^{Q}(t)$ is given by  Eq.~(\ref{h-01}). 

At zero temperature, we find \cite{Moskalets:2004ct}

\begin{eqnarray}
{\cal P}^{Q}_{\alpha\beta} =
\frac{1 }{2h }
\int   dE 
\sum\limits_{q=-\infty}^\infty  
\left\{ f_{0}\left(E_{} \right) - f_{0}\left(E_{q} \right) \right\}^{2}   
\nonumber \\
\sum\limits_{\gamma, \delta  = 1}^{N_r } 
\sum\limits_{n,m=-\infty}^\infty  
\left(E _{n} - \mu \right) 
\left(E _{m} - \mu  \right)
{\rm S} _{F,\alpha \gamma }^* \left( E _{n}, E \right)\, 
\label{hn-02} \\
{\rm S} _{F,\beta \gamma }^{} \left( E_{m}, E  \right)\,
{\rm S} _{F,\beta \delta }^* \left( E_{m} , E_{q}   \right) 
{\rm S} _{F,\alpha \delta }^{} \left( E _{n} , E_{q}   \right)\, 
. 
\nonumber
\end{eqnarray}

\subsection{Heat current cross-correlator}

We  make use of the unitarity of the Floquet scattering matrix, Eq.~(\ref{en-03}), and show that for $ \alpha \ne \beta$, only the term containing the product of the Fermi functions, $f_{0}(E) f_{0}(E_{q})$, does contribute to Eq.~(\ref{hn-02}). 
In the long-period driving limit introduced in Sec.~\ref{ldpl}, we get the heat cross-correlation noise,

\begin{eqnarray}
{\cal P}_{\alpha \ne \beta}^{Q}  =  
-\dfrac{1}{h^{2} {\cal T} }  
\sum\limits_{n,m=-\infty}^\infty  
\left(E _{n} - \mu \right) 
\left(E _{m} - \mu  \right)
\nonumber \\
\int dE f_{0}\left(E_{} \right)  
\sum\limits_{\gamma = 1}^{N_r } 
{\rm S} _{F,\alpha \gamma }^* \left( E _{n}, E \right)\, 
{\rm S} _{F,\beta \gamma }^{} \left( E_{m}, E  \right)\,
\label{hn-03} \\
\int dE_{q} f_{0}\left(E_{q} \right) 
\sum\limits_{\delta  = 1}^{N_r } 
{\rm S} _{F,\alpha \delta }^{} \left( E _{n} , E_{q}   \right)
{\rm S} _{F,\beta \delta }^* \left( E_{m} , E_{q}   \right) 
\, . 
\nonumber
\end{eqnarray}
\ \\ \noindent
Comparing this equation and Eq.~(\ref{CM-06}) for $ \alpha \ne \beta$, we relate the heat current cross-correlator to the elements of the excess first-order correlation matrix,

\begin{eqnarray}
{\cal P}_{\alpha \ne \beta}^{Q} &=& 
- \frac{ v_{ \mu}^{2} }{ {\cal T} }
\int _{0}^{ {\cal T} } dt_{1} 
\int _{ - \infty}^{ \infty } d (t_{2} - t_{1} )
\label{hn-04} \\
&&
\left( - i \hbar  \frac{\partial  }{\partial t_{1} } - \mu \right)G^{(1)}_{out, \alpha \beta }( t_{1};t_{2}) 
\nonumber \\
&& \times
\left( - i \hbar  \frac{\partial  }{\partial t_{2} } - \mu \right)G^{(1)}_{out, \beta \alpha  }( t_{2};t_{1}) .
\nonumber 
\end{eqnarray}
\ \\ \noindent 
To show explicitly that this quantity is real, one needs to make a set of transformations on the right hand side of this equation: Integrating by parts over both times and using Eq.~(\ref{sym-1}). 
Then, we arrive at an equation which is the complex conjugate of  Eq.~(\ref{hn-04}).  
Since this equation and its complex conjugate are the same, the equation in question is real. 

Note that Eqs.~(\ref{en-09}) and (\ref{hn-04}) remain also valid at finite temperatures when the outgoing leads $\alpha$ and $\beta$ are not directly connected, \textit{i.e.}  $S_{F, \alpha \beta} = 0$.
In this case, the thermal noise does not contribute to the current correlation function.\cite{Moskalets:2004ct}


\section{Ballistic electronic network}

\label{appf}

Few assumptions are required to enable analytical calculations concerning 
the mesoscopic conductor 
in question.
We first suppose that 
the conductor can be viewed as consisting of 
nodes, quantum point contacts, which are connected via chiral waveguides.  
The scattering process at these nodes is energy-independent.
Second, the input and output leads, $\gamma$ and $\alpha$, are connected via $N_{ \alpha \gamma}$ paths, that is,  

\begin{eqnarray}
S_{C, \alpha \gamma}(E) = 
\sum\limits_{ \ell=1}^{N_{ \alpha \gamma}} 
S_{ \alpha \gamma}^{ \ell} e^{i\left( k(E) L_{ \alpha \gamma}^{ \ell} +  \varphi_{ \alpha \gamma}^{ \ell} \right)} ,
\label{CM-09}
\end{eqnarray}
\ \\ \noindent
where $k(E) L_{ \alpha \gamma}^{ \ell}$ is a kinematic phase accumulated by an electron with energy $E$ along the trajectory $ \ell$ connecting the input $ \gamma$ and the output $ \alpha$ and $ \varphi_{ \alpha \gamma}^{ \ell}$ is a corresponding phase due to possibly present magnetic flux.  
Note that the coefficients $S_{ \alpha \gamma}^{ \ell}$ are energy-independent. 
We name such a conductor {\it a ballistic electronic network}.

To simplify again the notation, we introduce the scattering amplitude, $S_{in, \gamma}(t,E)$ whose Fourier coefficients define the Floquet scattering matrix of the source in the lead $ \gamma$ as follows,  \cite{Moskalets:2008ii} 

\begin{eqnarray}
S_{F, \gamma}(E_{n},E) = S_{in, \gamma, n}(E) \equiv \int _{0}^{ {\cal T} } \frac{dt }{ {\cal T} } e^{in \Omega t} S_{in, \gamma}(t,E) .
\nonumber \\
\label{CM-10}
\end{eqnarray}
\ \\ \noindent
With all assumptions and transformations made, the excess electronic correlation matrix elements, Eq.~(\ref{CM-06}), can be cast into the following form, 

\begin{eqnarray}
G^{(1)}_{out, \alpha \beta}( t_{1};t_{2}) = 
\frac{ 1 }{hv_{ \mu} }
\int   dE f_{0}(E)
e^{i   \left( t_{1} - t_{2} \right) \frac{ E }{\hbar} } 
\nonumber \\
\label{CM-11} \\
\Bigg\{
\sum_{ \gamma = 1 }^{ N_{r} } 
\sum\limits_{ \ell=1}^{N_{ \alpha \gamma}} 
\sum\limits_{ \ell^{\prime}=1}^{N_{ \beta \gamma}} 
e^{-i   \left( \tau_{ \alpha \gamma}^{ \ell} - \tau_{ \beta \gamma}^{ \ell^{\prime}} \right) \frac{ E }{\hbar} } 
e^{-i\left( \varphi_{ \alpha \gamma}^{ \ell} - \varphi_{ \beta \gamma}^{ \ell^{\prime}} \right)}
\left( S_{ \alpha \gamma}^{ \ell}  \right)^{*}
\nonumber \\
\times
S_{in, \gamma}^{*}( t_{1} - \tau_{ \alpha \gamma}^{ \ell}, E) 
S_{ \beta \gamma}^{ \ell^{\prime}} 
S_{in, \gamma}( t_{2 } - \tau_{ \beta \gamma}^{ \ell^{\prime}}, E)    
- 1 \Bigg\} .
\nonumber 
\end{eqnarray}
\ \\ \noindent
Here $ \tau_{ \alpha \gamma}^{ \ell} = L_{ \alpha \gamma}^{ \ell} / v_{ \mu}$. 
The scattering at the nodes is supposed to be instantaneous. 

As an illustration let us consider a simple but instructive example.

\subsection{Conductor with a single wave splitter} 

Let us consider two chiral waveguides connected to each other via a quantum point contact (QPC). 
Each incoming channel, $ \gamma = 1,2$, is fed by a single-electronic source described by the scattering matrix $S_{in, \gamma}$.  
The scattering matrix of the QPC is a $2\times 2$ unitary matrix. 
For the present purpose,  we choose it as follows, 

\begin{eqnarray}
{\bf S} = 
\left ( \begin{array}{ll} \sqrt{R}  & i\sqrt{T}  \\ \ \\ i\sqrt{T}  & \sqrt{R}  \end{array}  \right ) ,
\label{CM-12}
\end{eqnarray}
\ \\ \noindent
where $T$ and $R= 1- T$ are energy-independent transmission and reflection probabilities at the QPC, respectively. 

Since the 
conductor 
has only one node, the scattering matrix of the 
conductor 
is energy-independent, ${\bf S}_{C} \equiv {\bf S}$. 
In this case, the excess-correlation matrix for outgoing electrons becomes

\begin{eqnarray}
{\bf G}_{out}^{(1)}( t_{1};t_{2}) &=&  {\bf S}^{*}  {\bf G}_{in} ^{(1)}( t_{1};t_{2}) {\bf S}^{T} ,
\label{CM-13} 
\end{eqnarray}
\ \\ \noindent
where ${\bf G}_{in}^{(1)}$ is a diagonal matrix

\begin{eqnarray}
{\bf G}_{in} ^{(1)} &=&
\left ( \begin{array}{ll} G^{(1)}_{1} & 0 \\  0 & G^{(1)}_{2} \end{array} \right ), 
\label{CM-14}
\end{eqnarray}
\ \\ \noindent
describing an incoming state with no inter-channel correlations since the two sources are independent. 
The entry $G_{ \gamma}^{(1)}$,  $ \gamma = 1,2$ corresponds to the excess correlation functions and describes  the particles injected by an electronic source into the lead $ \gamma$, 

\begin{eqnarray}
G^{(1)}_{ \gamma}( t_{1};t_{2}) &=& 
\int   \frac{ d E    f\left( E \right) }{hv_{ \mu} }
e^{i   \left( t_{1} - t_{2} \right) \frac{ E }{\hbar} } 
\nonumber \\
\label{CM-15} \\
&& \times
\left\{
S_{in, \gamma}^{*}( t_{1}, E) 
S_{in, \gamma}( t_{2 }, E)    
- 1 \right\} .
\nonumber 
\end{eqnarray}
\ \\ \noindent

\subsubsection{The purity condition}

If each source emits particles in a pure state, then the corresponding correlation functions satisfy the following equation,

\begin{eqnarray}
v_{ \mu}\int _{- \infty}^{ \infty } dt G^{(1)}_{ \gamma}( t_{1};t) G^{(1)}_{ \gamma}( t;t_{2}) = G^{(1)}_{ \gamma}( t_{1};t_{2}) .
\label{CM-16}
\end{eqnarray}
\ \\ \noindent
In this case, it is easy to show that the outgoing state is also pure. 
Indeed,  the corresponding correlation matrix does satisfy the same (but a matrix) equation, 

\begin{eqnarray}
v_{ \mu}\int _{- \infty}^{ \infty } dt 
{\bf G}_{out}^{(1)}( t_{1};t)
{\bf G}_{out}^{(1)}( t;t_{2})
= 
{\bf G}_{out}^{(1)}( t_{1};t_{2}) .
\label{CM-17}
\end{eqnarray}

Note that, in general, the state projected onto any single outgoing lead becomes a mixed state unless: (i) the sources emit states  characterized by the same correlation functions and (ii) these states overlap perfectly at the QPC, i.e., $G^{(1)}_{ 1}( t_{1};t_{2}) = G^{(1)}_{ 2}( t_{1};t_{2})$ after the QPC. 
To make this clear, let us rewrite a matrix equation (\ref{CM-13}) in terms of its components,  

\begin{equation}
{\bf G}_{out}^{(1)} = 
\left ( 
\begin{array}{ll}
R G^{(1)}_{1} + T G^{(1)}_{2}
&
i\, \sqrt{RT} \left( G^{(1)}_{1} -  G^{(1)}_{2} \right)
\\ \ \\
 i\, \sqrt{RT} \left( G^{(1)}_{2} -  G^{(1)}_{1} \right)
&
T G^{(1)}_{1} + R G^{(1)}_{2}
\end{array}
 \right ) . 
\label{CM-18} 
\end{equation}
\ \\ \noindent 
The state emitted, say, into the outgoing lead $ \alpha=1$ is described by the following correlation function, 
${ G}_{out,11}^{(1)} = R G^{(1)}_{1} + T G^{(1)}_{2}$. 
If  $G^{(1)}_{ 1} \ne G^{(1)}_{ 2}$,  this state is a mixed state, composed of a state with a correlation function $G^{(1)}_{1}$ that appears with probability $R$ and a state with a correlation function $G^{(1)}_{2}$ that appears with probability $T$. 
However, if $G^{(1)}_{1} = G^{(1)}_{2} \equiv { G}_{}^{(1)}$, then the emitted state is a pure state, since ${ G}_{out,11}^{(1)} = { G}_{}^{(1)}$ (remember that the state described by ${ G}_{}^{(1)}$ is a pure state). 

Another interesting conclusion, which can be deduced from Eq.~(\ref{CM-18}), is the following. 
The non-diagonal elements of the correlation matrix for the outgoing state, ${ G}_{out,12}^{(1)}$, ${ G}_{out,21}^{(1)}$, depend only on the difference of correlation functions of the incoming particles, $G^{(1)}_{1} -  G^{(1)}_{2} $. 
Therefore, if the incoming states are composed of more than one particles, say,  $G^{(1)}_{1} = G^{(1)}_{s} - G^{(1)}_{a}$ and $G^{(1)}_{2} = G^{(1)}_{s} + G^{(1)}_{a}$, then the off-diagonal elements of ${\bf G}_{out}^{(1)}$ keep information on $G^{(1)}_{a}$ while they completely lose  information on $G^{(1)}_{s}$. 
In contrast, the diagonal elements $ G_{out, \alpha \alpha}^{(1)}$, (in the case of a symmetric QPC, $T=R=1/2$) keep information on $G^{(1)}_{s}$ while they lose information on  $G^{(1)}_{a}$. 
This linear property of the correlation functions can be used, for example, to separate out a single-particle contribution from the multi-particle one, see, e.g., Ref.~\cite{Moskalets:2016vt}.

\subsubsection{Single-particle incoming states}

\label{spis}

When the sources emit single particles in a pure state, we have  $G^{(1)}_{ \gamma}(t_{1}; t_{2}) = \Psi_{ \gamma }^{*}(t_{1}) \Psi_{ \gamma }(t_{2})$, $ \gamma = 1,2$. 
Using the non-diagonal elements of the matrix ${\bf G}_{out}^{(1)}$, Eq.~(\ref{CM-18}), and the normalization condition, $ \int_{- \infty}^{ \infty} d t \left | \Psi_{ \gamma}( t)  \right |^{2} = 1/ v_{ \mu}$, it is straightforward to show that Eqs.~(\ref{hn-04}) and (\ref{en-09}) are reduced to Eqs.~(\ref{hpn-1}) and (\ref{hpn04}), respectively.

%
%

\end{document}